\documentclass[twocolumn,prd,aps]{revtex4-1}
\usepackage{graphicx}
\usepackage{hyperref}
\usepackage{amssymb}
\usepackage{amsmath}
\usepackage{xcolor}
\usepackage{lineno}
\usepackage{enumitem}

\newcommand{\appropto}{\mathrel{\vcenter{
  \offinterlineskip\halign{\hfil$##$\cr
    \propto\cr\noalign{\kern2pt}\sim\cr\noalign{\kern-2pt}}}}}

\begin{document}

\title{Implications of Strong Intergalactic Magnetic Fields for Ultra-High-Energy Cosmic-Ray Astronomy}

\author{Rafael Alves Batista$^1$}
\email{rafael.alvesbatista@physics.ox.ac.uk}
\author{Min-Su Shin$^2$}
\author{Julien Devriendt$^1$}
\author{Dmitri Semikoz$^{3,4}$}
\author{Guenter Sigl$^5$}
\affiliation{
$^1$ Department of Physics - Astrophysics, University of Oxford, DWB, Keble Road, Oxford, OX1 3RH, United Kingdom \\
$^2$ Korea Astronomy and Space Science Institute, 305-348, Daejeon, Republic of Korea \\
$^3$ APC, Universite Paris Diderot, CNRS/IN2P3, CEA/IRFU, Observatoire de Paris, Sorbonne Paris Cite, 119 75205 Paris, France \\
$^4$ National Research Nuclear University MEPHI (Moscow Engineering Physics Institute), Kashirskoe highway 31, Moscow, 115409, Russia \\
$^5$ II. Institut f\"ur Theoretische Physik, Universit\"at Hamburg, Luruper Chaussee 149, 22761, Hamburg, Germany
}

\begin{abstract}
We study the propagation of ultra-high-energy cosmic rays in the magnetised cosmic web. We focus on the particular case of highly magnetised voids ($B \sim \text{nG}$), using the upper bounds from the Planck satellite. The cosmic web was obtained from purely magnetohydrodynamical cosmological simulations of structure formation considering different power spectra for the seed magnetic field in order to account for theoretical uncertainties. We investigate the impact of these uncertainties on the propagation of cosmic rays, showing that they can affect the measured spectrum and composition by up to $\simeq 80\%$ and $\simeq 5\%$, respectivelly. In our scenarios, even if magnetic fields in voids are strong, deflections of 50 EeV protons from sources closer than $\sim\;$50 Mpc are less than $15^\circ$ in approximately 10-50\% of the sky, depending on the distribution of sources and magnetic power spectrum.
Therefore, UHECR astronomy might be possible in a significant portion of the sky depending on the primordial magnetic power spectrum, provided that protons constitute a sizeable fraction of the observed UHECR flux.

\end{abstract}



\maketitle

\section{Introduction}

The origin of the ultra-high-energy cosmic rays (UHECRs), particles with energies above $\sim 1\; \text{EeV}$ ($1\;\text{EeV} = 10^{18}\;\text{eV}$), is an open problem in astrophysics. While the mechanisms of acceleration of UHECRs are not known, Fermi acceleration is widely regarded as one of the most likely scenarios, and the observation of cosmic rays correlating with specific types of astrophysical objects may enable us to uncover their sources.

Cosmic rays at ultra-high energies are predominantly atomic nuclei. Recent measurements by the Pierre Auger Observatory~\cite{auger2014a,auger2014b} are compatible with a light proton-dominated composition around $1\;\text{EeV}$, which becomes heavier as energy increases. The second largest experiment, the Telescope Array, reports a lighter composition at the highest energies~\cite{ta2015a}. Nevertheless, it has been shown that both composition estimates are consistent within uncertainties~\cite{auger2015a}.

The UHECR spectrum contains remarkable features whose exact interpretations are a matter of debate. At $E \approx 5\;\text{EeV}$ there is a change in the spectral index, the so-called ``ankle''. At $E \approx 50\;\text{EeV}$ a cutoff is observed, which has been first interpreted in terms of interactions of UHE protons with the cosmic microwave background (CMB), the Greisen-Zatsepin-Kuzmin (GZK) cutoff~\cite{greisen1966a,zatsepin1966a}. This cutoff may also be a signature of the maximal energy attainable by cosmic accelerators~\cite{aloisio2011a}.

Cosmic rays with energies $E\gtrsim 10\;\text{EeV}$ likely originate from extragalactic sources~\cite{hillas1967a,blumenthal1970a,berezinsky2004a,berezinsky2006a}. Results by the Pierre Auger Observatory~\cite{auger2013a} support the extragalactic origin of cosmic rays at the highest energies, although the presence of a galactic component at energies of a few EeV is not excluded and has been advocated by several authors (see e.g. refs.~\cite{allard2005a,aloisio2014a}).

The Telescope Array Collaboration (TA) has found indications of an intermediate-scale of $\sim 20^\circ$ centred at equatorial coordinates $(\alpha, \delta)=(146.7^\circ,43.2^\circ)$, for energies $E \gtrsim 57 \; \text{EeV}$~\cite{ta2014a}. Auger has also found an excess of UHECRs coming from within a window of $\sim 18^\circ$ around the nearest active galactic nuclei, Cen A, for a similar energy range~\cite{auger2014b}. The identification of the source (if it is unique) associated with these excesses, if they are not spurious statistical fluctuations, depends on the intervening magnetic fields, which combined with a predominantly heavy composition may explain such large angular spreads.

In order to interpret observations and look for the elusive sources of UHECRs, it is crucial to understand the propagation of these particles in the universe. Hence, a detailed understanding of the processes through which they lose energy and the intervening magnetic fields which cause their deflection is essential. 

The structure and strength of the galactic magnetic field (GMF) can be modelled by combining synchrotron maps with rotation measures of extragalactic sources, as done by Jansson and Farrar~\cite{jansson2012a,jansson2012b}. The extragalactic magnetic field (EGMF), however, is largely unknown, and its strength and structure are sources of uncertainties. There are several magnetohydrodynamical (MHD) simulations of the cosmic web (e.g. ~\cite{miniati2002a,dolag2004a,das2008a}). These simulations often predict different structure for the fields, and their strengths may differ by a few orders of magnitude from each other. Few observational constraints on EGMFs exist, making it more difficult to construct a model for them than for galactic fields.

Seminal works by Sigl {\it et al.}~\cite{sigl2003a,sigl2004a,sigl2004b} and Dolag {\it et al.}~\cite{dolag2004a,dolag2005a} have discussed the deflection of UHECRs in EGMFs and the prospects for identifying their sources making use of MHD simulations of the magnetised cosmic web. Simulations of the propagation of UHECRs performed by the former suggest high deflections ($\delta \sim 20^\circ$ for $100\;\text{EeV}$ protons), whereas the latter obtained small deflections ($\delta \lesssim 1^\circ$ at $100\;\text{EeV}$), which would make UHECR astronomy possible. Later works by Das {\it et al.}~\cite{das2008a} indicate moderate deflections, $\delta \lesssim 15^\circ$ at $60\;\text{EeV}$ in most of the sky. Kotera \& Lemoine~\cite{kotera2008a} have not explicitly estimated deflections, but based on the fact that the voids in their simulation are highly magnetised (nG-level), one can expect deflections to be large.  Recent work by Hackstein {\it et al.}~\cite{hackstein2016a} have shown that the nearby distribution of sources up to 50 Mpc dominates the anisotropy signal, enabling UHECR (proton) astronomy.

In light of the aforementioned discrepant results, in this work we study the propagation of UHECRs in the magnetised cosmic web assuming the most extreme case of magnetisation, namely voids with fields $\sim 1 \; \text{nG}$. This paper is structured as follows: in section~\ref{sec:MHD} we present the MHD simulations used in this work; in section~\ref{sec:simulationSetup} the setup of the simulations of UHECR propagation is described; section~\ref{sec:results} contains the results of the simulations accompanied by a detailed discussion; finally, in section~\ref{sec:summary} we summarise our findings.


\section{The simulated cosmic web}\label{sec:MHD}

Small fluctuations in the early universe can generate a cosmic web composed of voids, filaments, sheets and clusters. Based on a set of initial conditions, grid-based magnetohydrodynamical simulations can be solved with gravitational evolution, including the most relevant astrophysical processes. MHD simulations using RAMSES~\cite{teyssier2002a} were performed. RAMSES is a multiresolution adaptive mesh refinement code which allows higher resolution in regions where a certain refinement condition is satisfied.

The simulation code can be divided essentially in two parts. The dynamical core ensures the conservation of relevant quantities (e.g.~mass, momentum, and energy) and the condition $\vec{\nabla} \cdot \vec{B} = 0$, and at the same time provides a framework to solve ideal MHD. The second part is related to the parametrisations of physical processes such as source and sink terms, heating and cooling, chemical reactions, convection, dynamos, and feedback, among many others. We consider minimum subgrid models for star formation~\cite{rasera2006a} and radiative heating and cooling~\cite{haardt1996a}.

Four MHD simulations were done~\cite{shin2016a}, changing the initial conditions of the magnetic fields. The size of the comoving simulation volume is (200h$^{-1}$ Mpc)$^3$, where $h\approx0.7$ is the normalised Hubble constant.  The initial number of dark matter particles in the runs is $256^3$, and the resolution of each cell is refined up to 18 levels. This implies a minimum cell size of approximately $762 h^{-1} \; \text{pc}$. The fraction of the volume with dark matter haloes with $M > 10^{12} M_\odot$ is $\simeq 2 \times 10^{-4}$, with a maximum halo mass of $2.3 \times 10^{15} M_\odot$ at $z=0$. Resolution effects start to become relevant for volume fractions below $\sim 3.4 \times 10^{-3}$, affecting mostly the high-redshift and low-density regions of the mass function.
The details of the runs are described below:
\begin{itemize}[noitemsep,nolistsep]
	\item run F: fiducial run;
	\item run L: less magnetic power over small scales;
	\item run S: less magnetic power over large scales;
	\item run O: magnetic power only on large scales.
\end{itemize}
For faster access these runs were resampled onto uniform grids with $512^3$ cells at $z=0$. This resampling did not significantly affect the distribution of magnetic fields and densities in the simulation volumes.

\begin{figure*}
	\centering
	\includegraphics[width=\columnwidth]{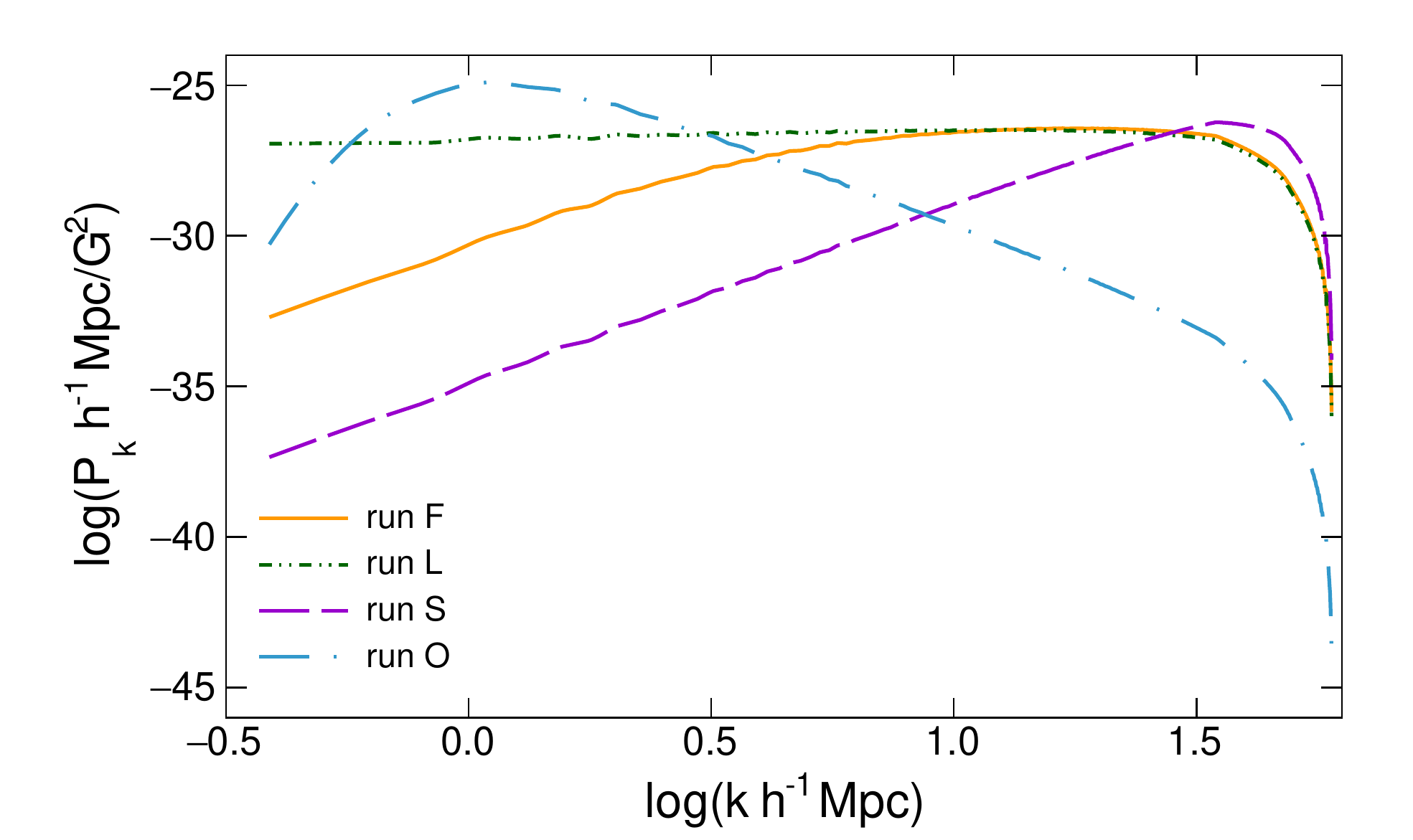}
	\includegraphics[width=\columnwidth]{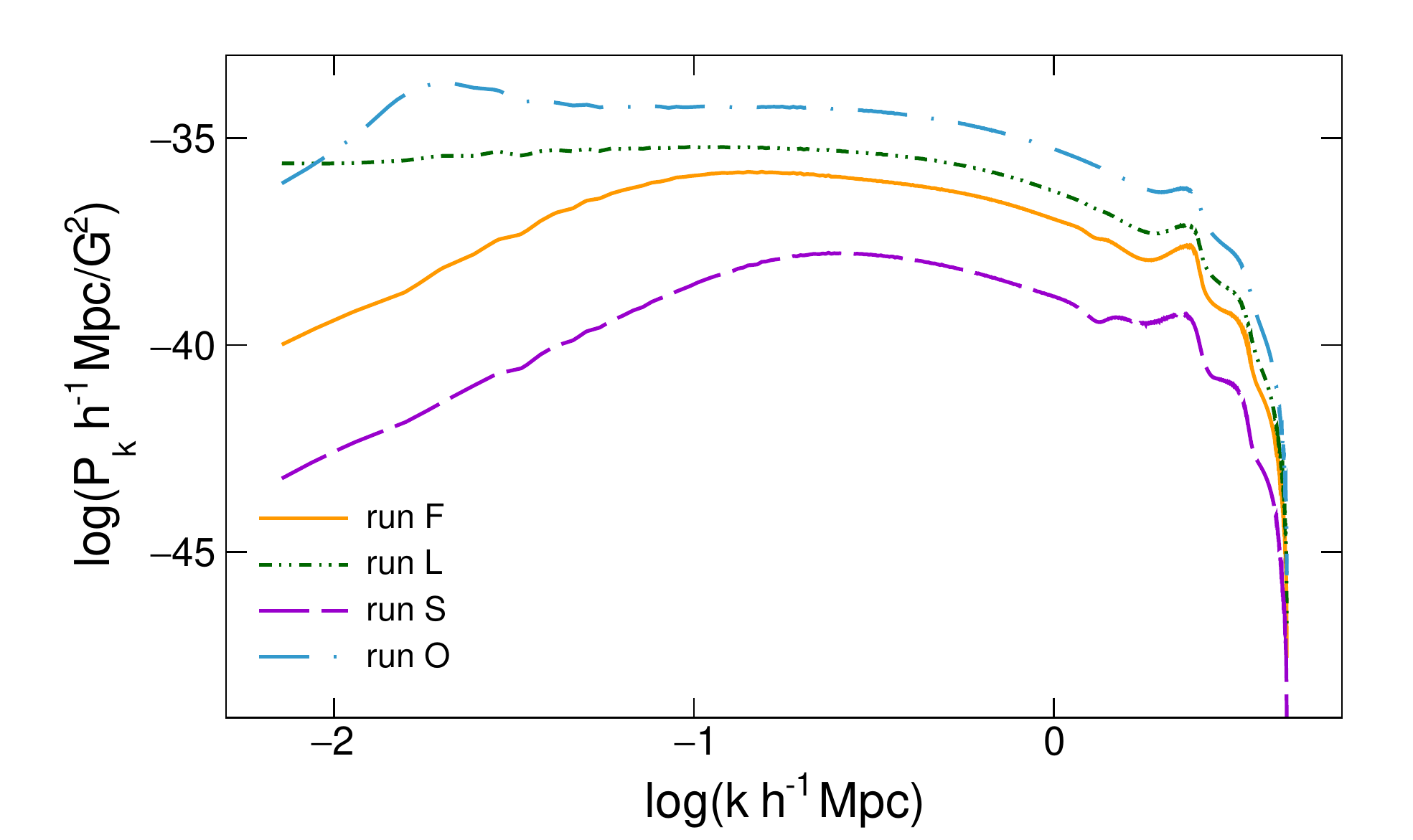}
	\caption{Power spectrum of the initial magnetic field at $z\simeq 53$ (left) and $z\approx 0$ (right panel) as a function of the comoving wave number $h^{-1}k$.}
	\label{fig:Bseed}
\end{figure*}

We have considered four different scenarios for the initial magnetic field in order to encompass theoretical uncertainties. These magnetic fields are described by gaussian random vector fields. Because the shape of the power spectrum of the seed field (injected at $z \approx 53$) affects the magnetic energy budget available during structure formation, it affects the distribution of magnetic field strengths today, and this could directly impact on the propagation of UHECRs. It is important to stress that in spite of the fact that the power spectra are different, the initial magnetic energy is the same in all runs. In Fig.~\ref{fig:Bseed} the power spectra of the magnetic field seeds are shown, together with their values at $z\approx0$. It is worth stressing that the initial conditions for the seed fields are not related to any specific model, and that they are so weak that the growth of large-scale structures is virtually unaffected by them. Moreover, the magnetic field and density perturbations are assumed to be uncorrelated, which might not be true.

The magnetic field strength can be scaled by an arbitrary factor without altering the distribution of magnetic fields in the simulation volumes, provided that the scaling factor does not exceed a given threshold~\cite{shin2016a}. Because of the uncertainties in the strength of magnetic fields in different regions of the universe, it is important to analyse limiting cases in order to put stringent bounds on the strength of EGMFs.


The choice of normalisations has to be consistent with observational constraints. Early studies using CMB data set an upper limit $B \lesssim 4\;\text{nG}$~\cite{barrow1997a} for intergalactic magnetic fields. The most recent results by the Planck Collaboration sets the upper bound to $4.4\;\text{nG}$ for zero helicity, and $5.6\;\text{nG}$ for maximal helicity, at comoving scales of 1 Mpc, at 95\% confidence level~\cite{planck2015a}. Magnetic fields in the centre of galaxy clusters can reach values as high as $B \sim 10\;\mu\text{G}$~\cite{bonafede2010a,vallee2011a}. Constraints based on Faraday rotation measurements using the NRAO VLA Sky Survey are $B \lesssim 1 \; \text{nG}$ for fields whose coherence lengths are of the order of the Jeans' length~\cite{pshirkov2016a}.

One can normalise magnetic field distributions using their filling factors, which indicate the fraction of the total volume filled with magnetic fields higher than a given reference value. Studies using data of the Sloan Digital Sky Survey (SDSS) suggest that the filling factor of voids is $\sim 0.2-0.9$~\cite{foster2009a,jasche2010a,pan2012a,leclercq2015a}, depending on how it is defined.

There are some indications that magnetic fields in filaments can reach up to $\sim 1\;\mu\text{G}$~\cite{ryu1998a,bagchi2002a}. Constraints from MHD simulations place the strength of these fields in the interval $\sim 1\;\text{nG}-1\;\mu\text{G}$~\cite{brueggen2005a,vallee2011a,vazza2014a,vazza2015a}. Analysis of SDSS data indicate that filaments fill about 1--10\%~\cite{forero-romero2009a,tempel2014a} of the volume.

A third bound for the choice of normalisation is the observed magnetic field in clusters of galaxies. Although this quantity is well-constrained by synchrotron and Faraday rotation measurements, the filling factors of clusters of galaxies are poorly known. Typical values for the strength of the field in these regions are $B \sim 0.05 - 30\;\mu\text{G}$ in the centre of clusters, and $\sim 10\;\text{nG}$ away from central regions, as suggested by synchrotron measurements of radio relics (see e.g. refs.~\cite{vallee2011a,widrow2002a} for reviews). Analyses of SDSS data~\cite{forero-romero2009a} indicate that the filling fraction of knots in the cosmic web are $f \sim 10^{-3}$. Nevertheless, one should bear in mind that the central regions of clusters, where magnetic fields are the highest, have much smaller volume filling factors than this, and that the magnetic field strength radially decreases away from the central regions of clusters.

Combined, these bounds indicate the regions of the parameter space corresponding to clusters, filaments, and voids, at $z=0$. Magnetic field strengths and volume filling factors of sheets are not well-known and therefore are not used as a constraint. We study here the case of strong intergalactic magnetic fields ($B \sim 1\;\text{nG}$) occupying about 80\% of the volume. It is important to mention that the normalisation adopted by us is only marginally consistent with measurements of magnetic fields in clusters of galaxies. As noted in Ref.~\cite{durrer2013a}, the efficiency of dynamos is uncertain and the effective amplification of magnetic fields may be as low as $10^3$, if dominated by compression during the collapse of structures. Nevertheless, because the voids and filaments occupy a volume much larger than clusters, the strength of the field in these regions plays a more important role than that in clusters, at least to leading order.

\begin{figure*}
	\includegraphics[width=2\columnwidth]{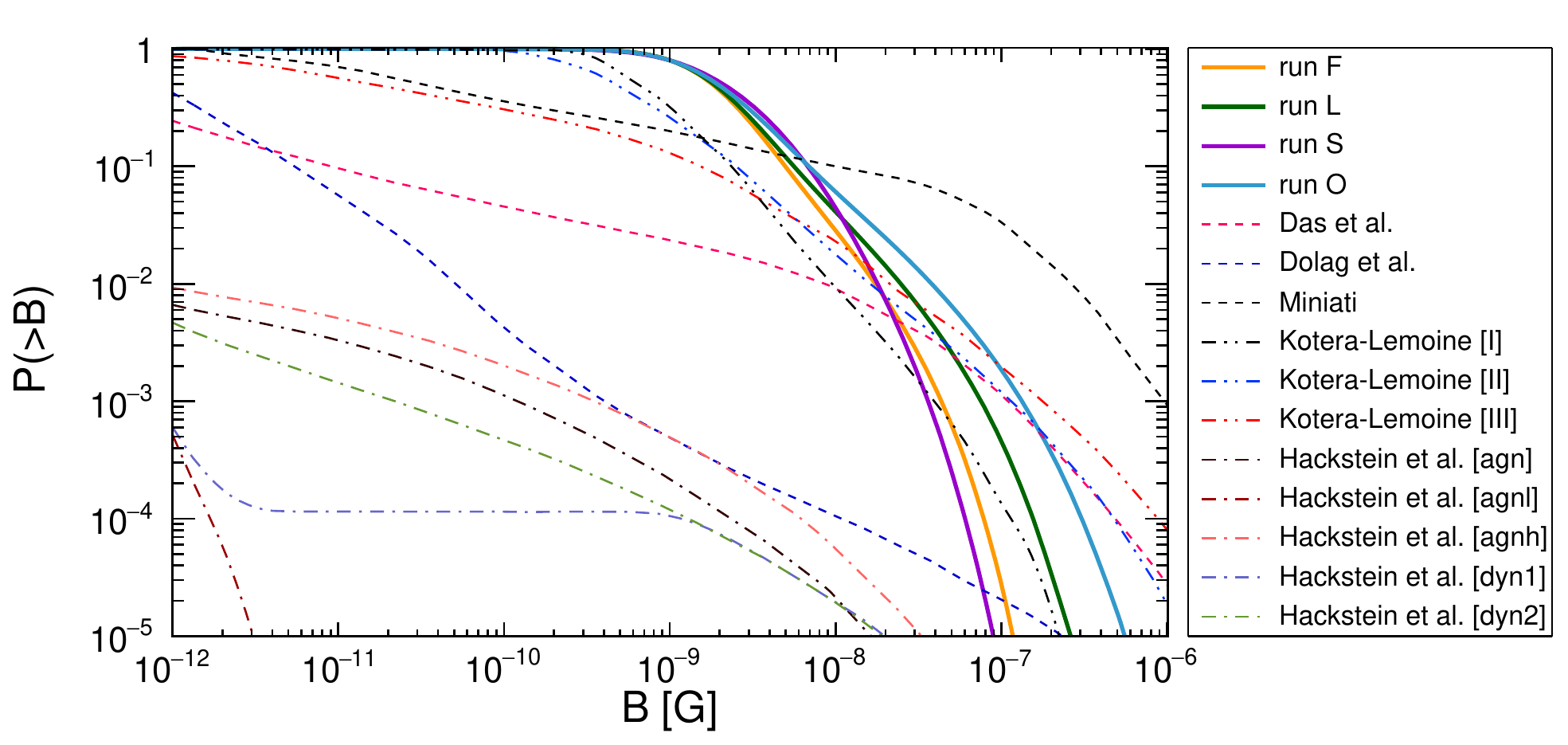}
	\caption{Cumulative filling factors for all the runs analysed in the present work (solid thick lines), together with a few models found in the literature~\cite{miniati2002a,dolag2004a,sigl2004a,das2008a,kotera2008a,hackstein2016a}.}
	\label{fig:fillingFactors}
\end{figure*}

The filling factors for our models, together with other models used in the literature, are presented in Fig.~\ref{fig:fillingFactors}. 

Dolag {\it et al.}~\cite{dolag2004a,dolag2005a} used a uniform magnetic field seed with strength $B \sim 1 \;\text{nG}$ at $z \approx 20$ to perform a constrained simulation of the local universe. In spite of the fact that different normalisations for this seed were studied and the final one chosen in such a way as to match observations of clusters, the effects of non-uniform seeds were not analysed, even though this is acknowledged as a potential source of uncertainty~\cite{dolag2005a}. The authors argue that a uniform magnetic field would result in the largest deflections for a given magnetic field strength, and therefore this could be taken as an upper limit. In our work we have used gaussian random vector fields as seed magnetic fields. 

In the work of Sigl {\it et al.}~\cite{sigl2003a,sigl2004a}, based on previous work by Miniati~\cite{miniati2002a}, magnetic fields are generated via the Biermann battery~\cite{biermann1950a}. According to Ref.~\cite{sigl2004b}, even if a uniform seed is used instead of Biermman battery, the outcome of the MHD simulations are qualitatively the same, and discrepancies are likely due to different numerical methods used in the MHD simulations rather than different initial conditions. 

The model by Das {\it et al.}~\cite{das2008a}, on the other hand, uses the method presented in Ref.~\cite{ryu2008a} to infer the magnetic field energy, based on the vorticity and kinetic energy of the turbulent flow in the simulations.

 Kotera \& Lemoine~\cite{kotera2008a} have performed hydrodynamical simulations and then derived the magnetic field based on specific assumptions for the relationship between the magnetic field $B$ and the gas density $\rho_g$. Models I, II, and III presented in Fig.~\ref{fig:fillingFactors} are, respectively: $B \propto \rho_g^{2/3}$ (isotropic collapse), $B \propto \rho_g^{0.9}$ (some kinds of anisotropic collapse), and $B \propto \rho_g  [1 + (\rho_g / \bar{\rho}_g)^{-2}]$ (unmagnetised voids).

 In the recent models by Hackstein {\it et al.}~\cite{hackstein2016a} the magnetic field strength is renormalised a posteriori to the thermal energy of the gas. The models `dyn1' and `dyn2' contain dynamos operating in all overdense regions, and only within virialised haloes, respectively. The models designated by `agn', `agnl', and `agnh' account for AGN feedback with energies $10^{58} \; \text{erg}$, $10^{57} \; \text{erg}$, and $10^{59} \; \text{erg}$, respectively. Note that in these models the void fields are much weaker than in our case.

 By comparing all the different filling factors in Fig.~\ref{fig:fillingFactors} it is clear that the uncertainties in the filling factors of extragalactic magnetic fields are high, and the models differ not only in their overall normalisation, but also on the slopes. Moreover, for the same magnetic field strength, these models may differ by a few orders of magnitude, which shows that no single MHD simulation of the cosmic web can be deemed as completely realistic as there are many theoretical uncertainties and model dependencies. Nevertheless, some common behaviours can be noted accross the simulations. Model I ($B \propto \rho_g^{2/3}$) by Kotera \& Lemoine~\cite{kotera2008a} and model `agn' by Hackstein {\it et al.}~\cite{hackstein2016a}, for instance, have an overall similar slope to our models (runs F, L, S, and O). This, however, is not surprising since this behaviour is expected from the simplest case of ideal MHD in which $B \appropto \rho_g^{2/3}$; the main model dependence here, besides the overall normalisation of the field, which would shift the corresponding curves in Fig.~\ref{fig:fillingFactors} toward higher or lower magnetic fields, is the prescription for star formation. It is also important to stress that, as can be seen in models `agnl' and `agnh' by Hackstein {\it et al.}, the total energy released by AGNs affects mainly the small fraction of volume corresponding to the region where strong magnetic fields appear.


\section{Simulation setup}\label{sec:simulationSetup}

Runs F, S, L, and O were used for the propagation of UHECRs. Because the  exact position of the observer within these simulation grids is not known, i.e., the simulations are not constrained, the ambiguity induced by this choice is removed by considering a large number of observers distributed in the simulation volume, and averaging over them. Hence the obtained results reflect the average behaviour of the magnetic field distribution in these cosmological simulations.

The propagation was done using the CRPropa 3 code~\cite{alvesbatista2016a}. Particles are injected by sources with energies between $1\;\text{EeV}$ and $1000\;\text{EeV}$, with the following spectrum
\begin{equation}
	\dfrac{dN}{dE} \propto 
	\begin{cases}
	E^{-\alpha}  &  \text{if } E_{max} > E \\[1em]
	E^{-\alpha}  \exp \left( 1 - \dfrac{E}{E_{max}} \right) &  \text{if } E_{max} \leq E \\[1em]
	\end{cases},
	\label{eq:spec}
\end{equation}
where $\alpha_{\text{Fe}} = 1$ and $\alpha_{\text{p}} = 2$ are the spectral indices for the injected iron and proton scenarios, respectively. Here $E_{max}$ is the maximal energy. In this work we use $E_{max,\text{p}} = 500\;\text{EeV}$ for protons and $E_{max,\text{Fe}} = 156\;\text{EeV}$ for iron primaries. One should note that these choices are arbitrary.

The energy loss processes taken into account were photopion production, pair production, and photodisintegration (including nuclear decays) in the case of nuclei. The ambient photon fields considered were the cosmic microwave background (CMB) and the extragalactic background light (EBL). We have adopted the EBL model by Gilmore {\it et al.}~\cite{gilmore2012a}. Energy losses caused by the adiabatic expansion of the universe as well as the redshift evolution of the CMB and EBL densities were not taken into account in this approach as these effects are subdominant at $E \gtrsim 10\;\text{EeV}$, as argued in Ref.~\cite{alvesbatista2014b}.

Particles are propagated until one of the following break conditions is met: they reach the observer, considered to be a sphere of radius $R_{obs}=1\;\text{Mpc}$; their energy drops below the minimum energy threshold, set to $E_{min} = 1\;\text{EeV}$; or the propagation time of the cosmic rays is larger than a predefined value, $T_{max} = 4000\;\text{Mpc}/c$. The size of the observer was assumed to be $R_{obs}=1\;\text{Mpc}$, which is not realistic as it would be virtually a point. We do not expect the choice of $R_{obs}$ to significantly affect the results since any misleading effects arising from the finite observer size would be washed out when averaging over several observers.

We have considered two configurations for the distribution of sources. In the first, sources follow the baryon density obtained from the MHD simulation. In the second, sources are randomly placed in the simulation volume and their positions are drawn from a uniform distribution. The latter scenario is presented in order to understand the impact of an unstructured source distribution on the studied observables, but is not an accurate description of reality. More details about the impact of the source distribution on the deflection of UHECRs can be found in Ref.~\cite{sigl2004a}.

\section{Results and Discussion}\label{sec:results}

\subsection{Spectrum and composition}

\begin{figure*}
	\centering
	\includegraphics[width=\columnwidth]{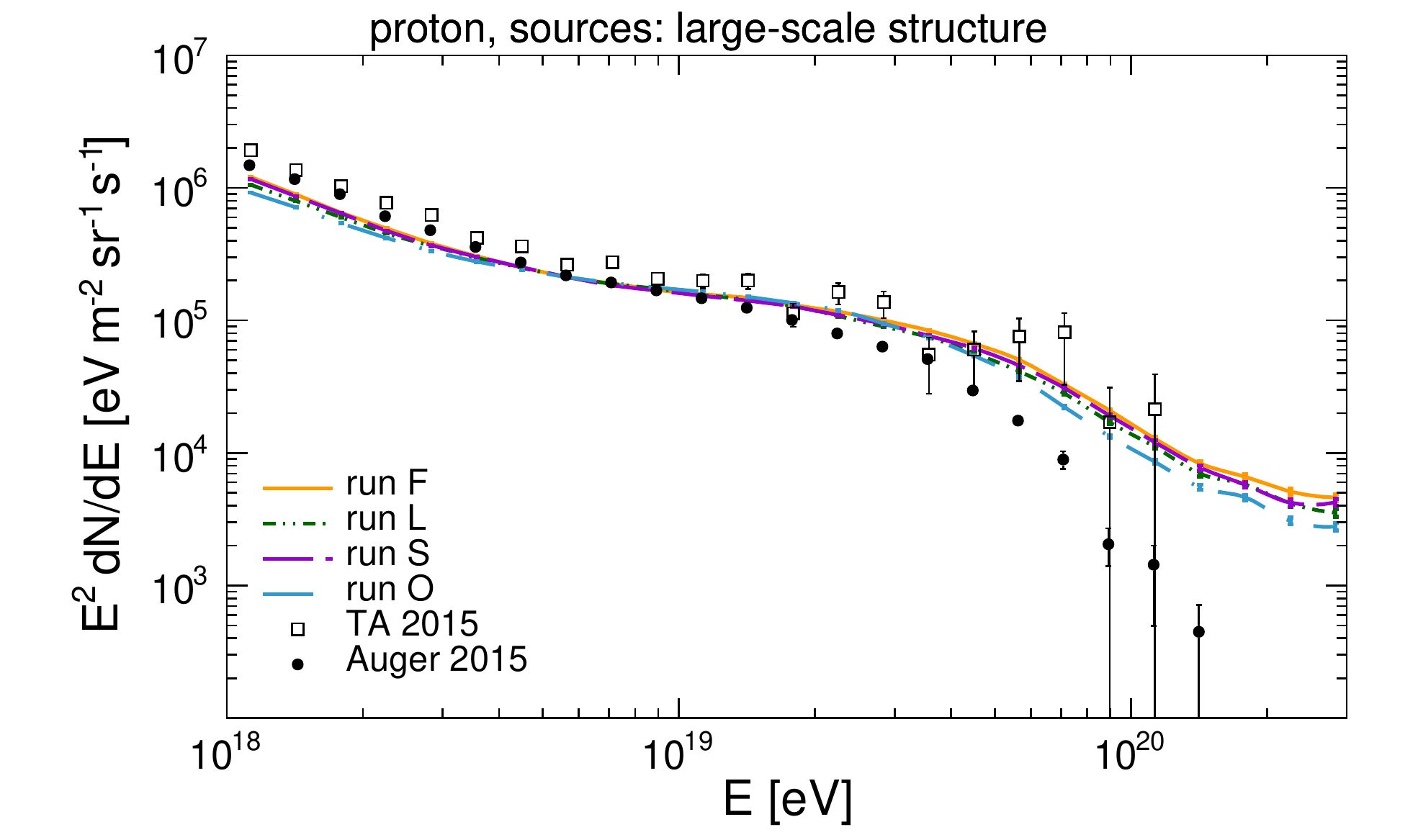}
	\includegraphics[width=\columnwidth]{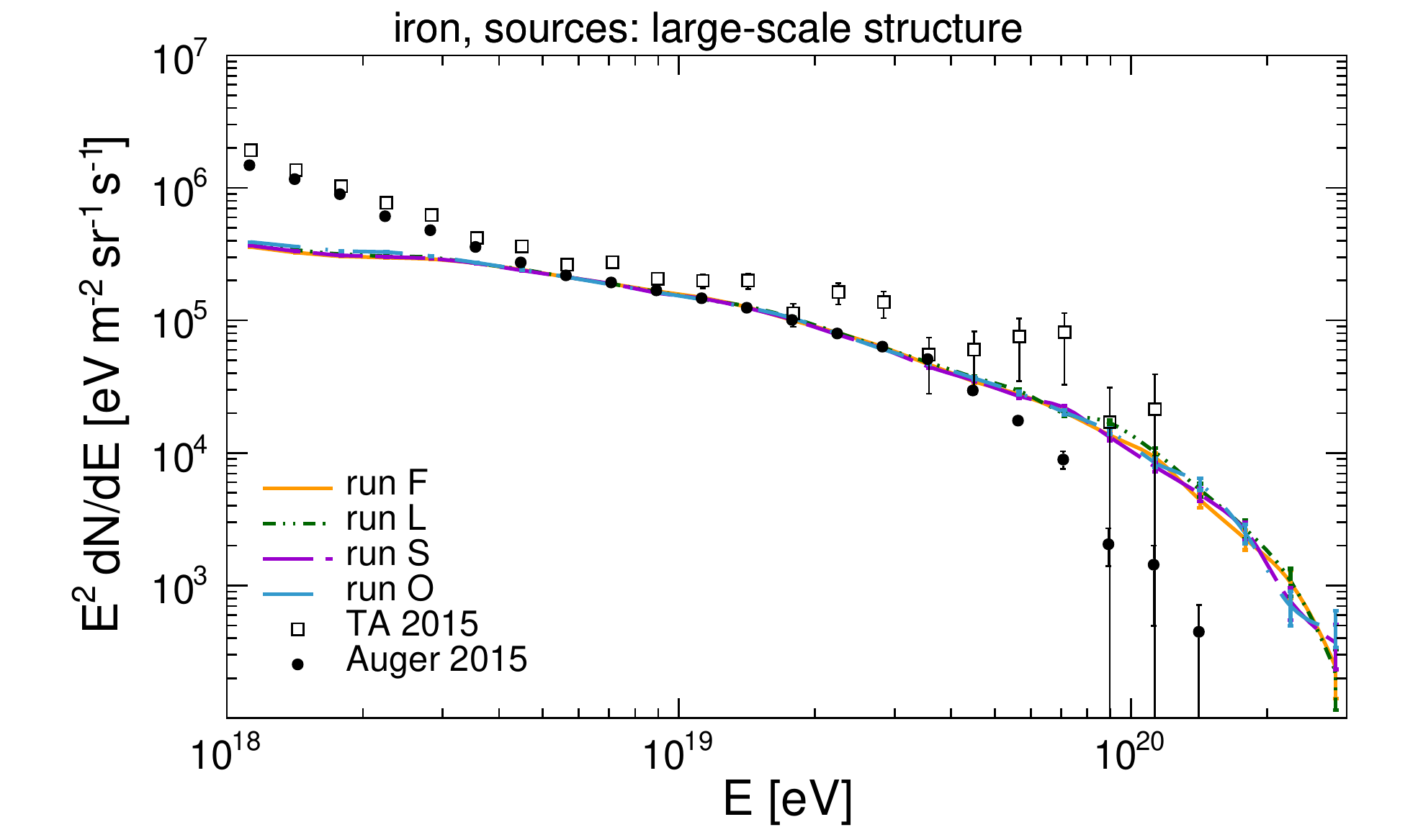}
	\caption{All-particle spectra for the different runs, assuming that sources follow the large-scale distribution of matter. The injected composition is assumed to be purely proton (left panel) and iron (right panel). Measurements by Auger~\cite{auger2015c} (circles) and TA~\cite{tinyakov2014a} (squares) are shown for comparison. Curves are arbitrarily normalised to Auger data at $E = 6 \; \text{EeV}$. Spectral parameters are, for the case of pure proton injection, $\alpha_\text{p} = 2$ and $E_{max,\text{p}}=500 \; \text{EeV}$, and $\alpha_\text{Fe}$ and $E_{max,\text{Fe}}=156 \; \text{EeV}$ for the pure iron injection scenario.}
	\label{fig:spec}
\end{figure*}

\begin{figure*}
	\centering
	\includegraphics[width=\columnwidth]{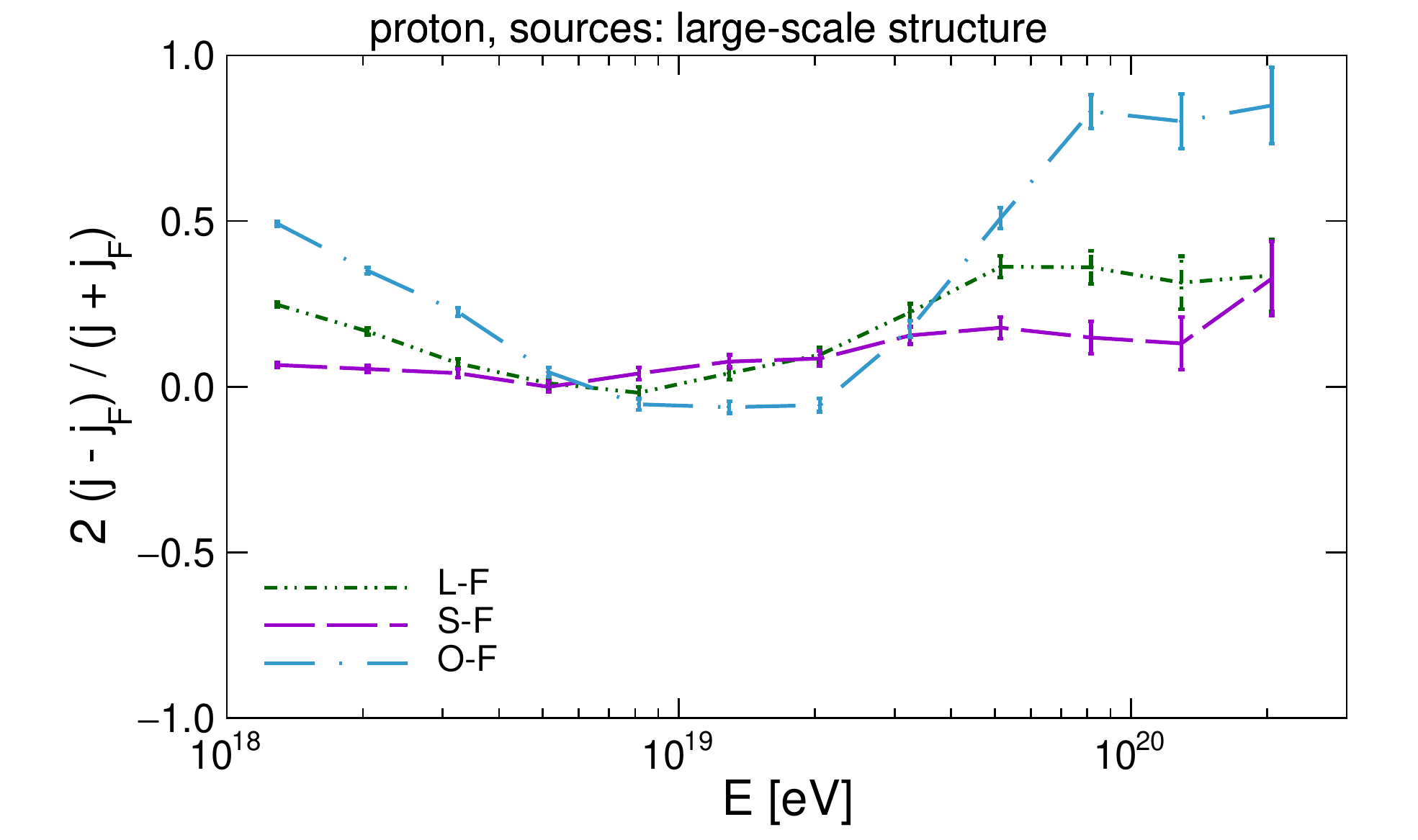}
	\includegraphics[width=\columnwidth]{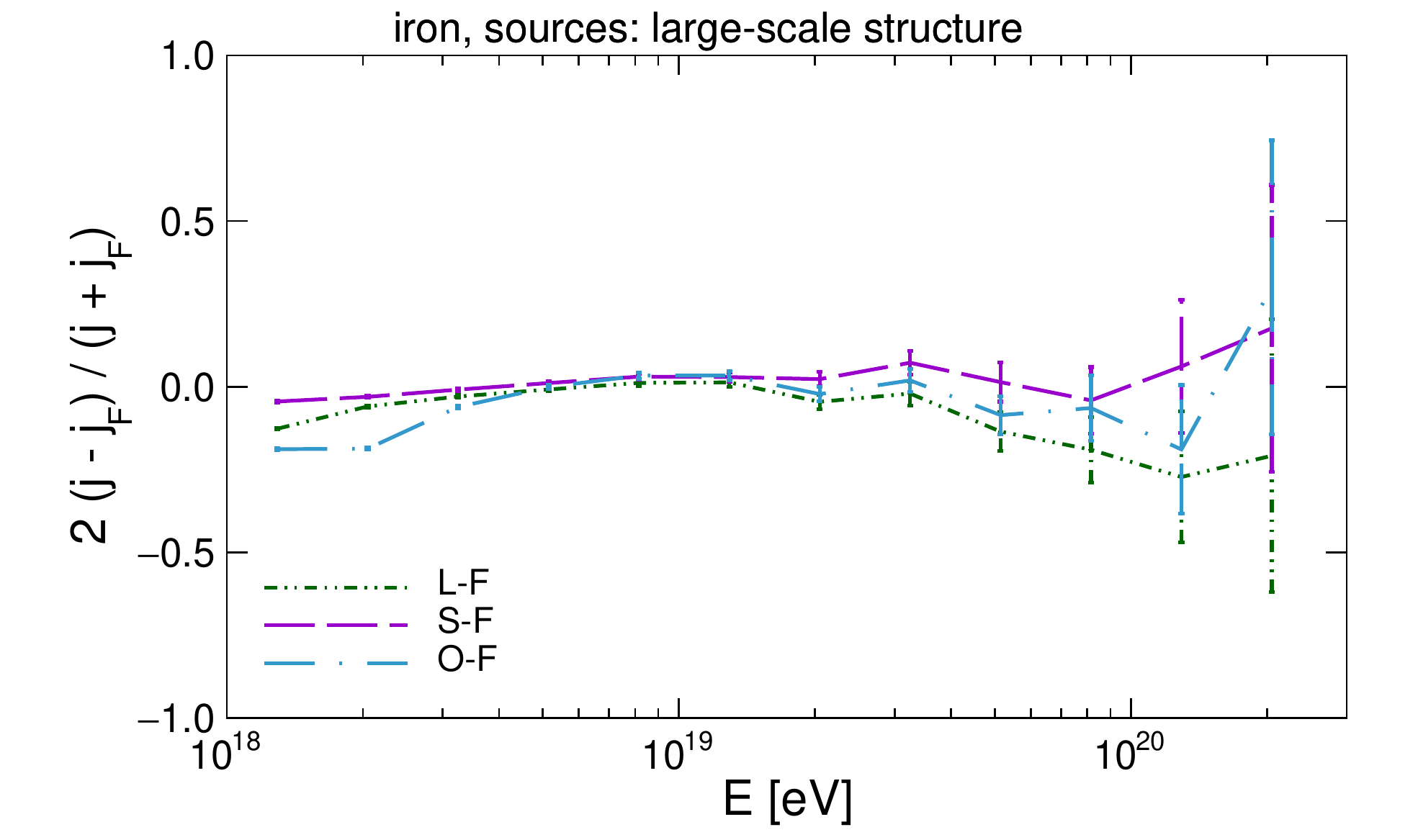}
	\caption{Difference between the spectra for the different runs, assuming that sources follow the large-scale distribution of matter. The injected composition is assumed to be purely proton (left) and iron (right panel). Spectral parameters are the same as in Fig.~\ref{fig:spec}.}
	\label{fig:specDiff}
\end{figure*}

In Fig.~\ref{fig:spec} we present the spectra obtained from the propagation of UHECRs in runs F, L, S, and O. The relative difference between runs L, S, and O, with respect to run F, are shown in Fig.~\ref{fig:specDiff}. One can see that the shape of the power spectrum of the magnetic field seeds leads to different spectral shapes. For reference, we present the spectra together with data measured by the Pierre Auger Observatory~\cite{auger2015c} and the Telescope Array~\cite{tinyakov2014a}. Note that our models are not realistic in terms of spectral properties (spectral indices, maximum energy, etc) nor composition, hence one should not expect our spectra to resemble the measurements. 

Firstly, it is notorious that the spectra for protons (Figs.~\ref{fig:spec} and~\ref{fig:specDiff}, left panels) are affected by the different intervening magnetic fields. These discrepancies are more pronounced at higher energies due to the arbitrarily chosen normalisation at $E = 6 \; \text{EeV}$.
Because our scenario is rather extreme, relevant propagation lengths such as the energy loss and diffusion lengths are comparable to the typical separation between sources, thus implying a non-universal spectrum as the propagation theorem~\cite{aloisio2004a} is not valid in this case. The propagation theorem states that for identical sources uniformly distributed within a given volume separated by distances much smaller than characteristic propagation quantities (e.g. diffusion length, energy loss length, etc), the diffusive spectrum of UHECRs has a universal form that does not depend on the modes of propagation. Note that this theorem holds for rectilinear propagation and diffusive propagation in weak and strong fields. Deviations from universality are expected, particularly at higher energies, due to inhomogeneities in the source distribution, flux enhancement/decrease due to the presence/absence of local sources, etc. The differences in the proton spectra for sources following the large-scale distribution and uniformly distributed are small because we have averaged over multiple observers to account for cosmic variance. In reality, if we had a constrained cosmological simulation with a single observer in a fixed position, the high-energy part of the spectrum would likely be affected by the presence or absence of nearby sources, making this difference more pronounced.

Secondly, the spectra for the iron scenario (Figs.~\ref{fig:spec} and~\ref{fig:specDiff}, right panels) are fairly similar for all magnetic field configurations studied, and compatible to each other within uncertainties. This can be understood considering that even at the highest energies deflections of UHE iron nuclei are high and directional information is almost completely lost. In this situation cosmic rays propagate a distance much larger than their Larmor radii before reaching the observer, if they arrive at the observer at all, smearing out effects of inhomogeneities. Averaging this effect over the nearest sources, it is reasonable to expect that the spectra for primary iron nuclei in these scenarios will not be sensitive to the magnetic field models considered since deflections are very high and effects of inhomogeneities are washed out. For lower energies, interaction horizons allow us to observe sources distant up to $\sim 1 \;\text{Gpc}$, energy range at which the flux is dominated by secondary nuclei stemming from the photodisintegration of primary iron nuclei, and consequently the effects of inhomogeneities are washed out by diffusive propagation.

In Fig.~\ref{fig:specDiff}, for any two spectra normalised for the same flux at $6 \; \text{EeV}$, the difference between them at any energy is at most 50\%.

 The observed spectral features will obviously depend on the properties of the accelerator such as $E_{max}$, $\alpha$, the cutoff shape, etc. 
We have used in this example an spectral index $\alpha_\text{p} = 2$ and $\alpha_\text{Fe} = 1$ for the purposes of illustration and discussion. The exact value of $\alpha$ depends on the acceleration mechanism, which is unknown. Furthermore, different sources may have different spectral indices, which would also affect the observed spectrum, particularly if the spectral indices of possible nearby sources which potentially dominate the flux deviate significantly from the average value of all sources.

\begin{figure*}[htb]
	\includegraphics[width=\columnwidth]{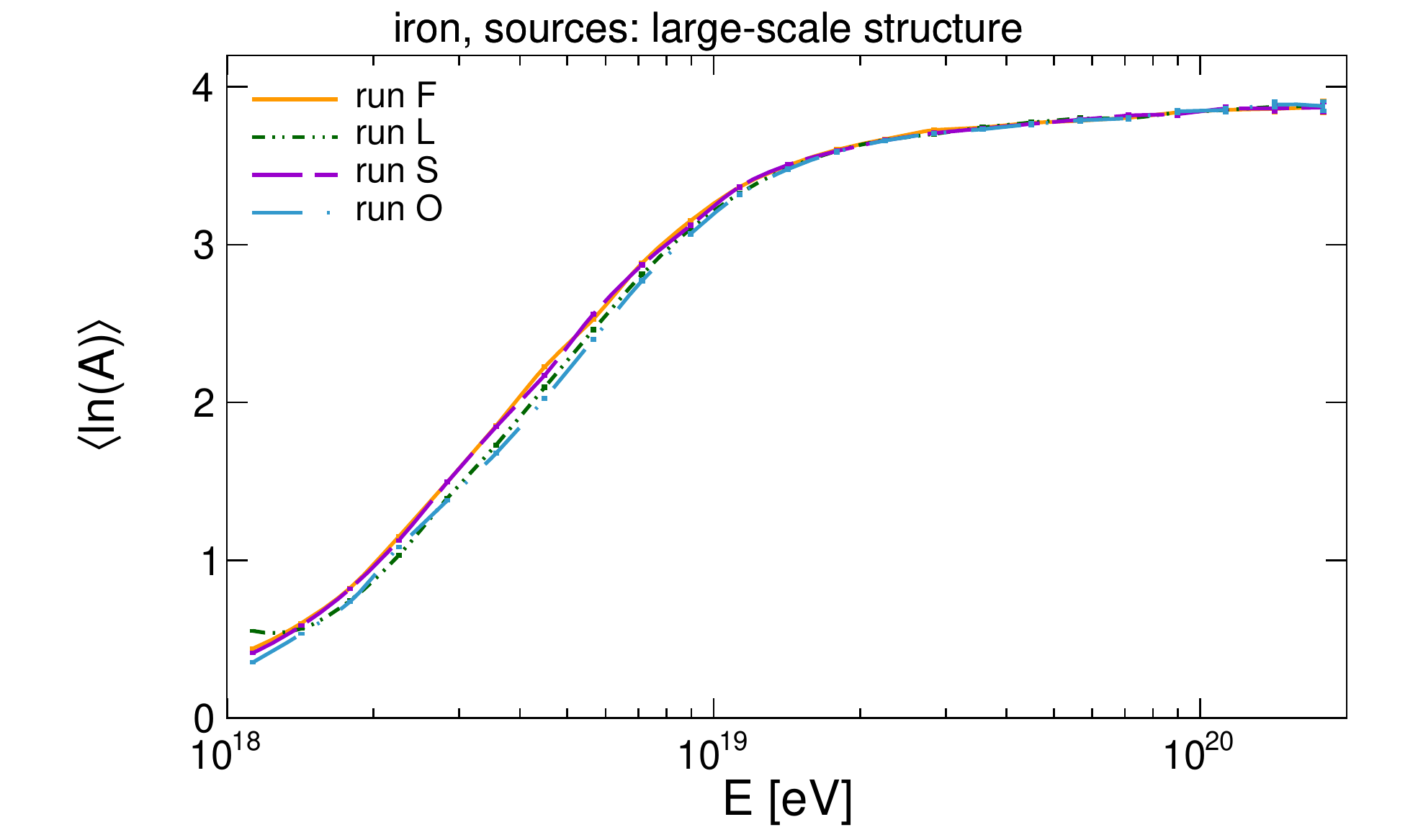}
	\includegraphics[width=\columnwidth]{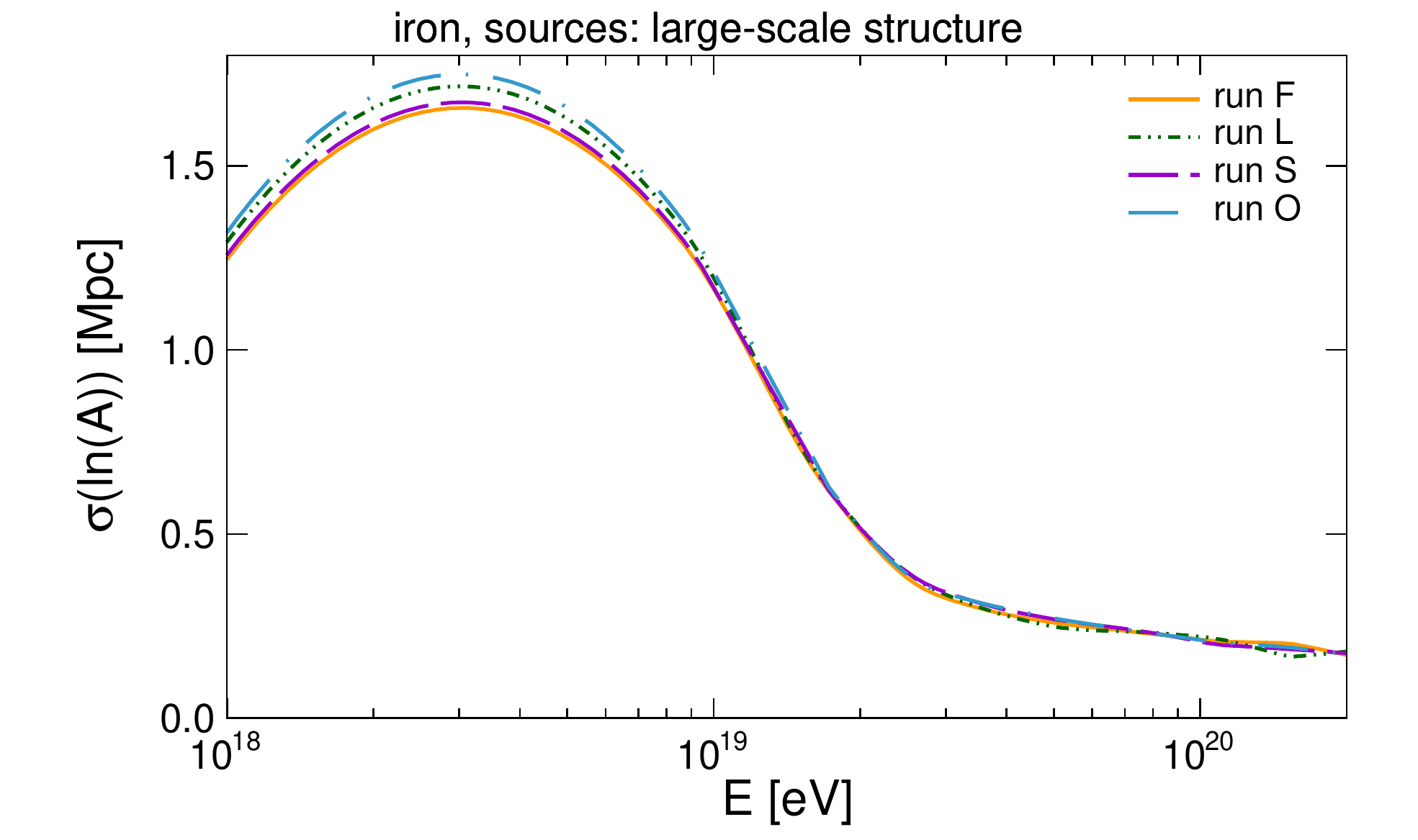}
	\caption{Average of the logarithm of the mass number ($\langle \ln A\rangle$, left panel) and its standard deviation ($\sigma(\ln A)$, right panel) for the case of pure iron injection assuming sources following the large-scale distribution of matter. Spectral parameters are the same as in Fig.~\ref{fig:spec}.}
	\label{fig:comp}
\end{figure*}

We have also compared the effects of the different runs on the composition of UHECRs measured at Earth. In Fig.~\ref{fig:comp} the average of the logarithm of the mass number $\langle \ln A\rangle$ and its standard deviation $\sigma(\ln A)$ are shown, for the injected iron scenario. Neither $\langle \ln A \rangle$ nor $\sigma(\ln A)$ differ significantly for the different runs. However, $\sigma(\ln A)$ is slighlty larger ($\sim 10\%$) for runs O and L compared to runs F and S in the energy range between 1 and 5 EeV.

The shape of the spectra depends also on the EBL model adopted, as shown in Ref.~\cite{alvesbatista2015a}. The differences are the largest in the energy range between $10$ and $50\;\text{EeV}$, which is roughly the same energy range at which we observe differences in the proton spectra for different runs. Therefore, if the goal were to constrain magnetic fields using UHECRs, it would be hard to discriminate between these sources of uncertainties.

\subsection{Magnetic horizon effects}

The spectrum may also be affected by the existence of magnetic horizons~\cite{lemoine2005a,aloisio2005a}. The magnetic horizon is defined as the maximum distance that cosmic rays can propagate away from their source for a given magnetic field configuration. In other words, if the time ($T$) it takes for a cosmic ray to propagate from its source and reach Earth is of the order of a Hubble time ($t_H \equiv H_0^{-1}$), i.e.~$T\sim t_H$, then a significant fraction of the flux of particles from this source will not be detected, thus causing a suppression in the flux of cosmic rays, which will be larger the lower the energy of the particles. 

The relevance of this effect at the highest energies ($E \gtrsim 10\;\text{EeV}$) is not clear. Studies assuming simple configurations of magnetic fields with fixed r.m.s. strengths were done in Refs.~\cite{globus2008a,mollerach2013a,alvesbatista2014a}, and the results suggest that it starts to become relevant at $E \lesssim 1\;\text{EeV}$ for $B \gtrsim 1\; \text{nG}$ and coherence lengths $l_c \sim 0.1-1\;\text{Mpc}$. More detailed studies~\cite{kotera2008a,alvesbatista2014a} considering a distribution of magnetic fields in the cosmic web suggest that the energy below which the flux might be significantly suppressed due to magnetic horizons is $\sim 0.1-1\;\text{EeV}$. 

\begin{figure*}
	\includegraphics[width=\columnwidth]{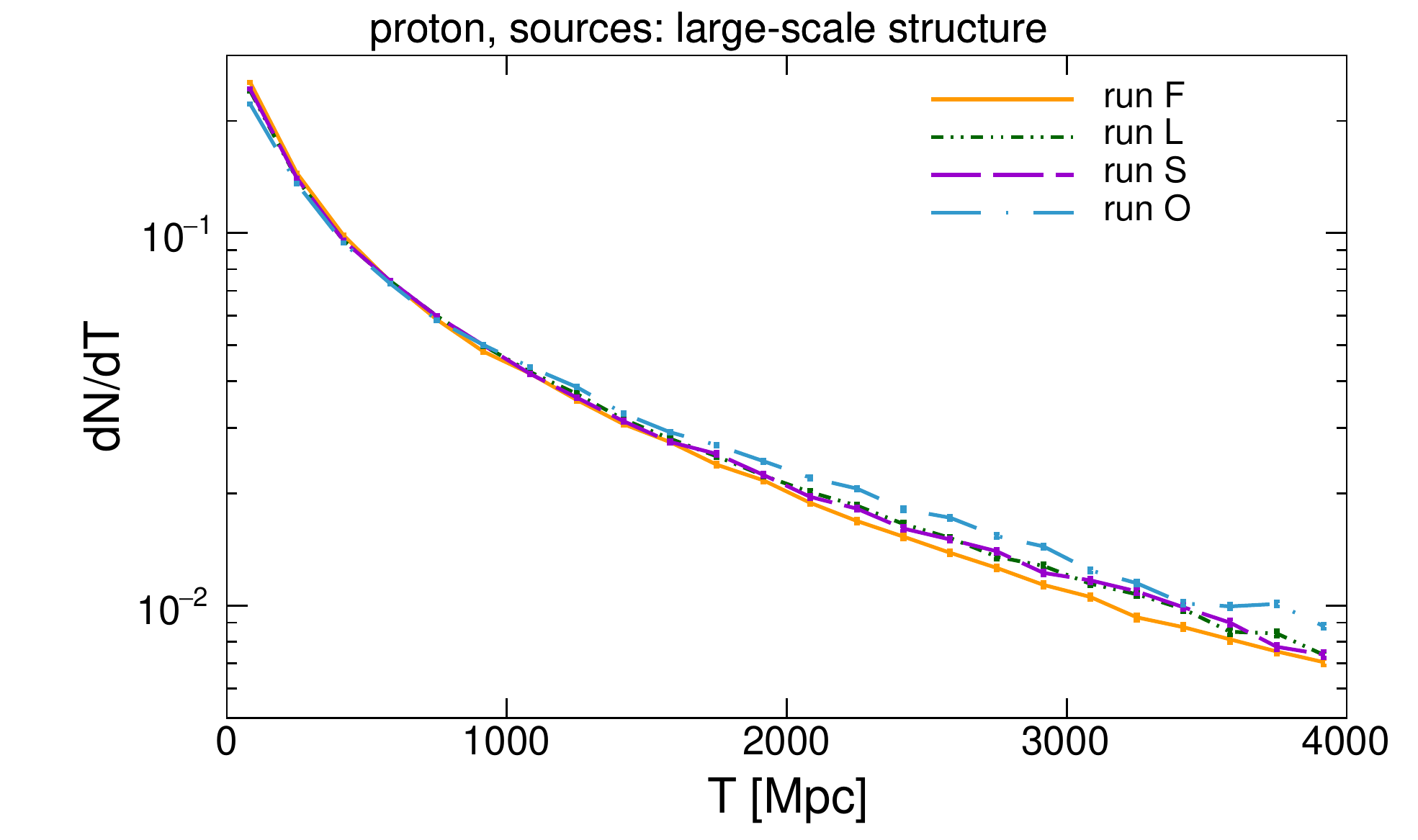}
	\includegraphics[width=\columnwidth]{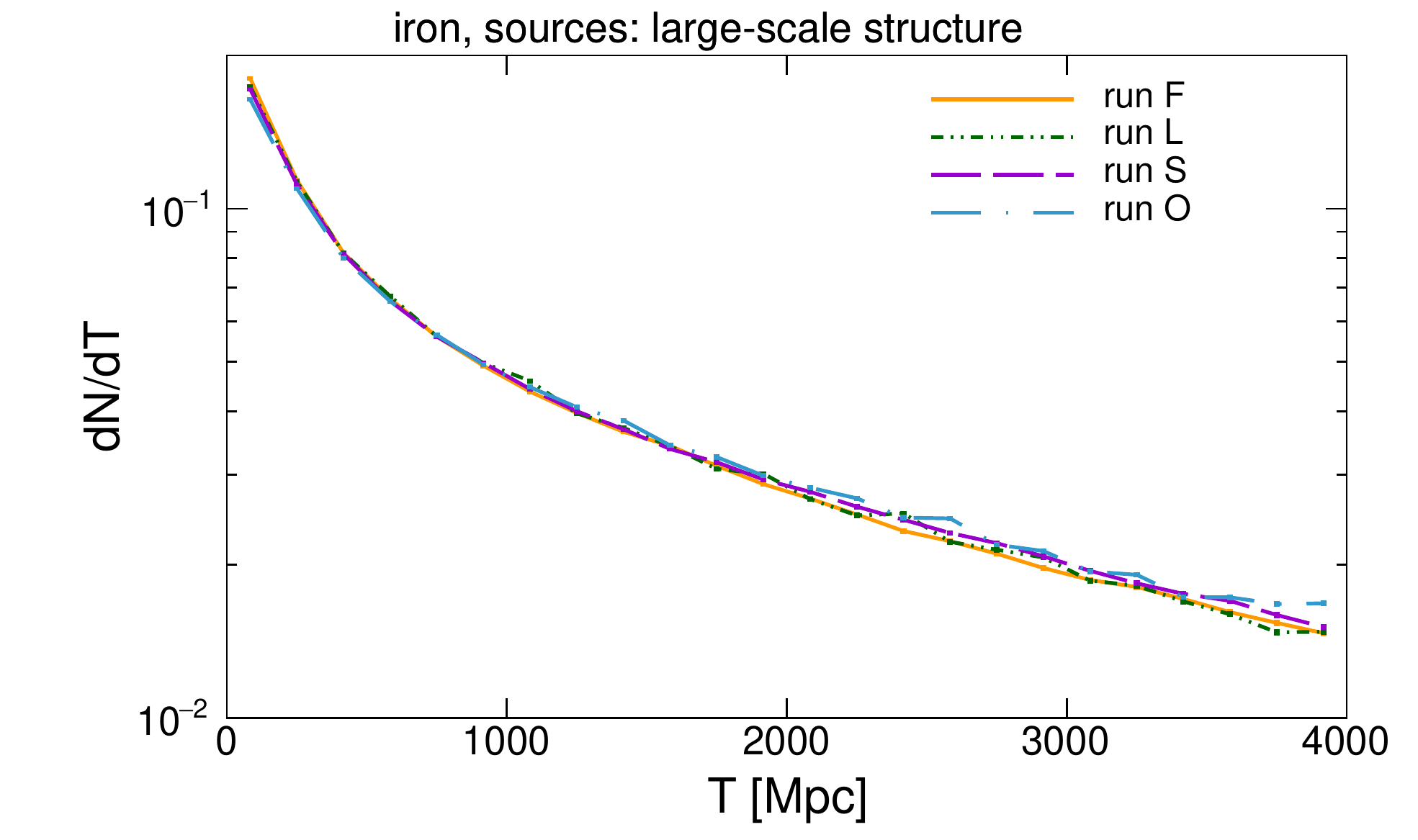}
	\caption{Distribution of propagation times for the different runs, assuming that sources follow the large-scale distribution of matter. The composition is assumed to be purely proton (left) and iron (right panel). Spectral parameters are the same as in Fig.~\ref{fig:spec}.}
	\label{fig:time}
\end{figure*}

As shown in Fig.~\ref{fig:time}, in this work we observe that a fraction ($\sim 10^{-2}$) of the total number of events have propagation times comparable to a Hubble time, i.e.~$T  \sim t_H$, suggesting the existence of a magnetic horizon at energies below a few EeV. This effect is, as expected, more prominent in the iron scenario. 

The verification of the existence of a magnetic horizon at $E \sim 1\;\text{EeV}$ has important implications for understanding the transition between galactic and extragalactic cosmic rays~\cite{thoudam2016a}, because it could, for example, cause the suppression of the flux of protons for $E \lesssim 0.1\;\text{EeV}$, which is consistent with many models~\cite{lemoine2005a,kotera2008a,globus2008a}.

\subsection{Deflections}

\begin{figure*}
	\centering
	\includegraphics[width=\columnwidth]{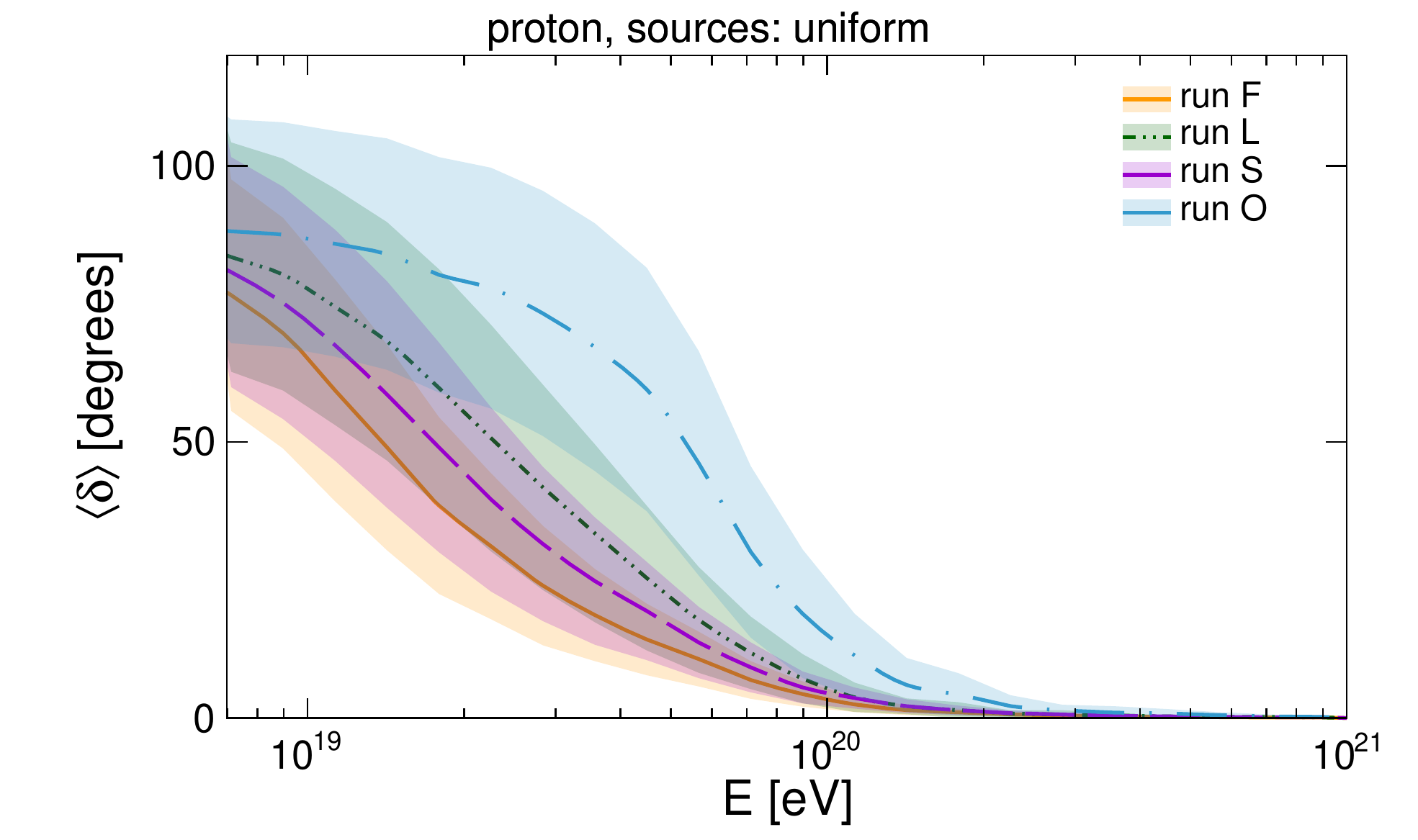}
	\includegraphics[width=\columnwidth]{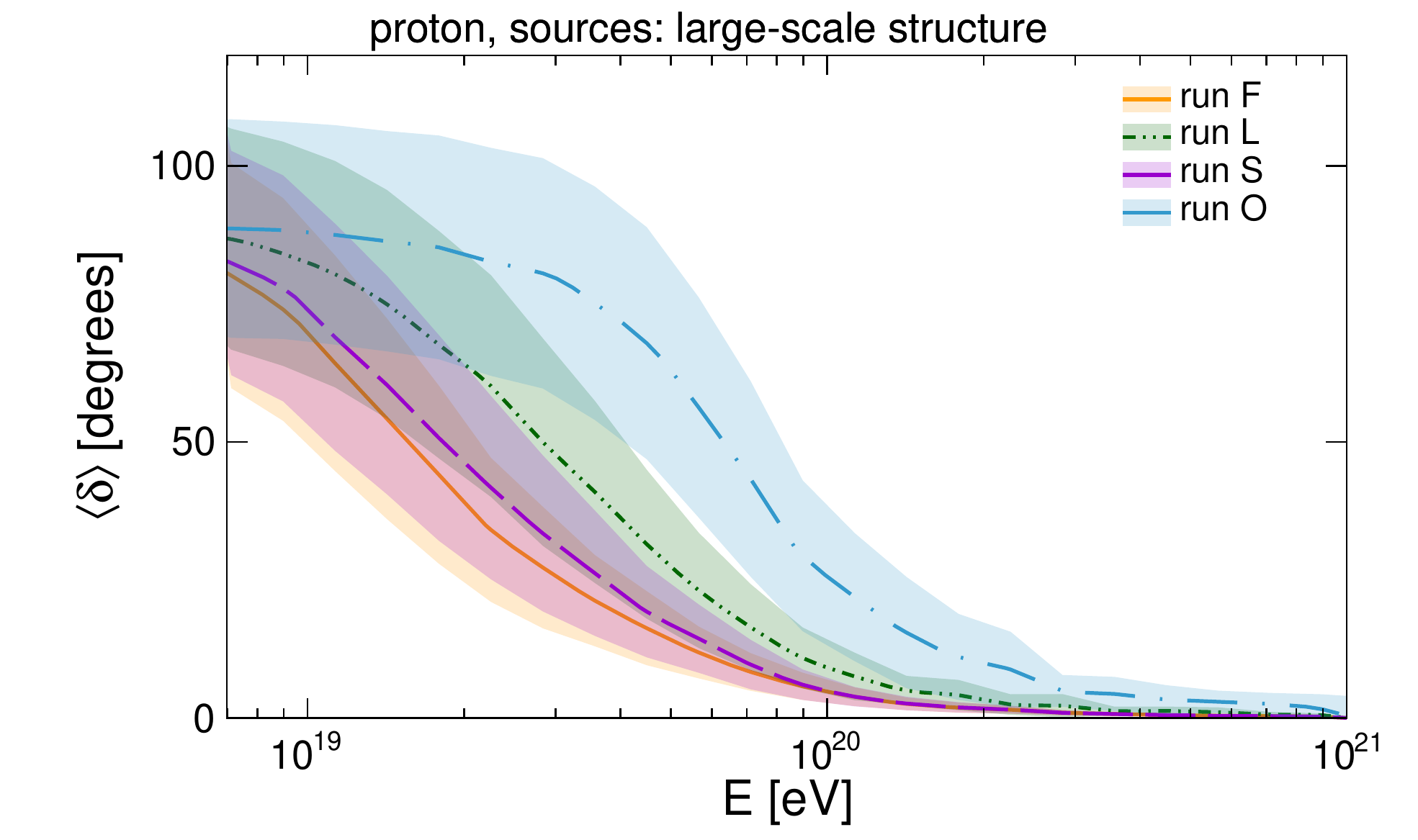}
	\caption{Average deflection as a function of the energy, and their corresponding standard deviations (hatched regions) for uniformly distributed sources (left) and sources following the simulated baryon density (right panel), emitting protons with the spectrum given by Eq.~\ref{eq:spec}. Spectral parameters are the same as in Fig.~\ref{fig:spec}.}
	\label{fig:deflE_Scaling_p}
\end{figure*}

In Fig.~\ref{fig:deflE_Scaling_p} the average deflection $\langle \delta \rangle$ of protons as a function of the energy is shown. This is calculated by considering all detected particles in a given energy bin, averaged over all observers. For a source distribution following the large-scale structure distribution (right panel), deflections are considerably larger than for the case of a uniform distribution (left panel), showing that the distribution of sources directly affects the estimated deflections. 

One can also notice in Fig.~\ref{fig:deflE_Scaling_p} that as the energy decreases deflections increase, converging to $\langle \delta \rangle \approx 90^\circ$ at $E \lesssim 10^{19} \; \text{eV}$, energy below which most of the flux is roughly isotropised. The exact value of this energy threshold depends on the source distribution -- for sources uniformly distributed this energy is lower than for sources following the large-scale distribution of matter.

Deflections for the iron scenario at $E \approx 100 \; \text{EeV}$ are much higher ($\langle \delta \rangle \gtrsim 50^\circ$) and are omitted.

\begin{figure*}
	\centering
	\includegraphics[width=\columnwidth]{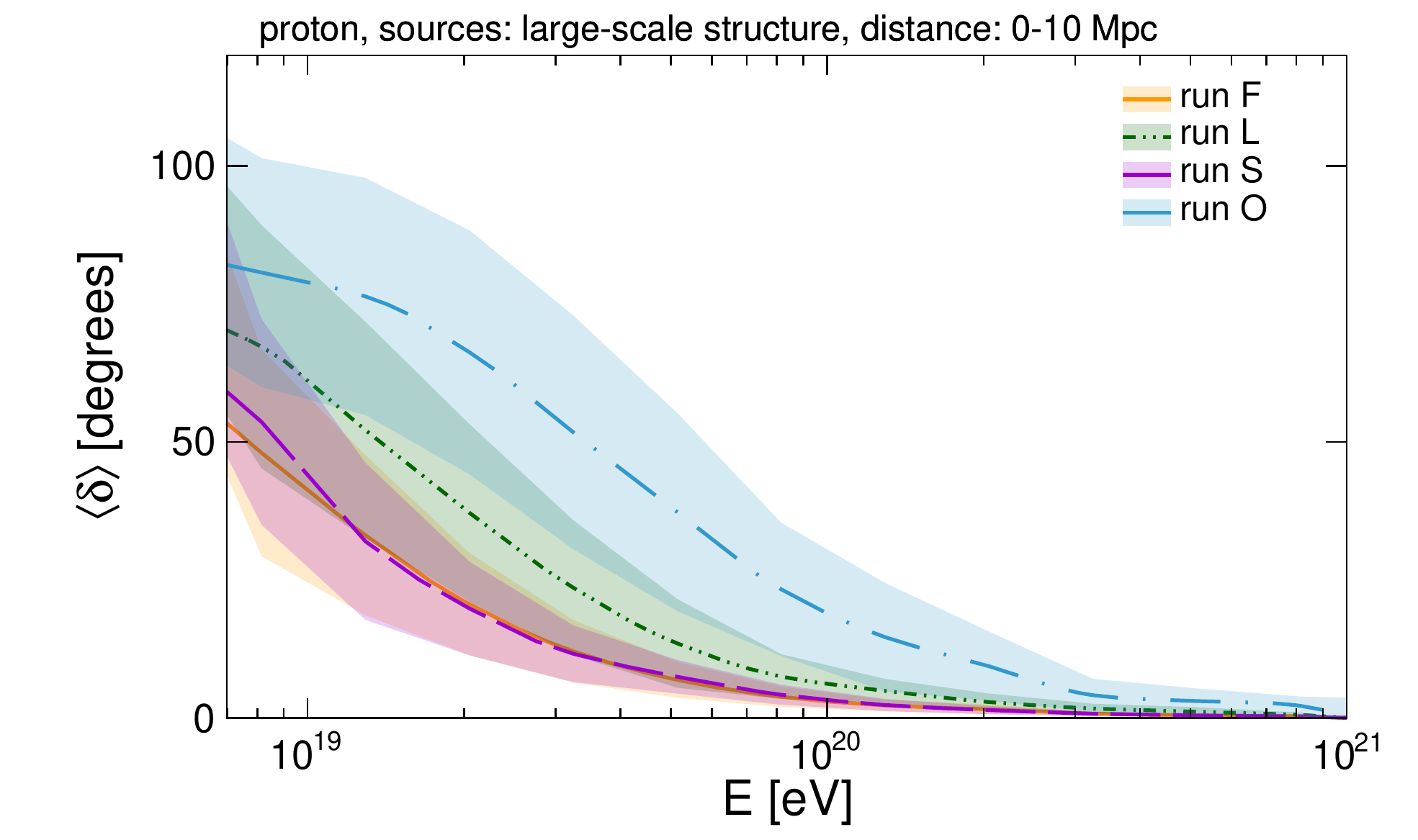}
	\includegraphics[width=\columnwidth]{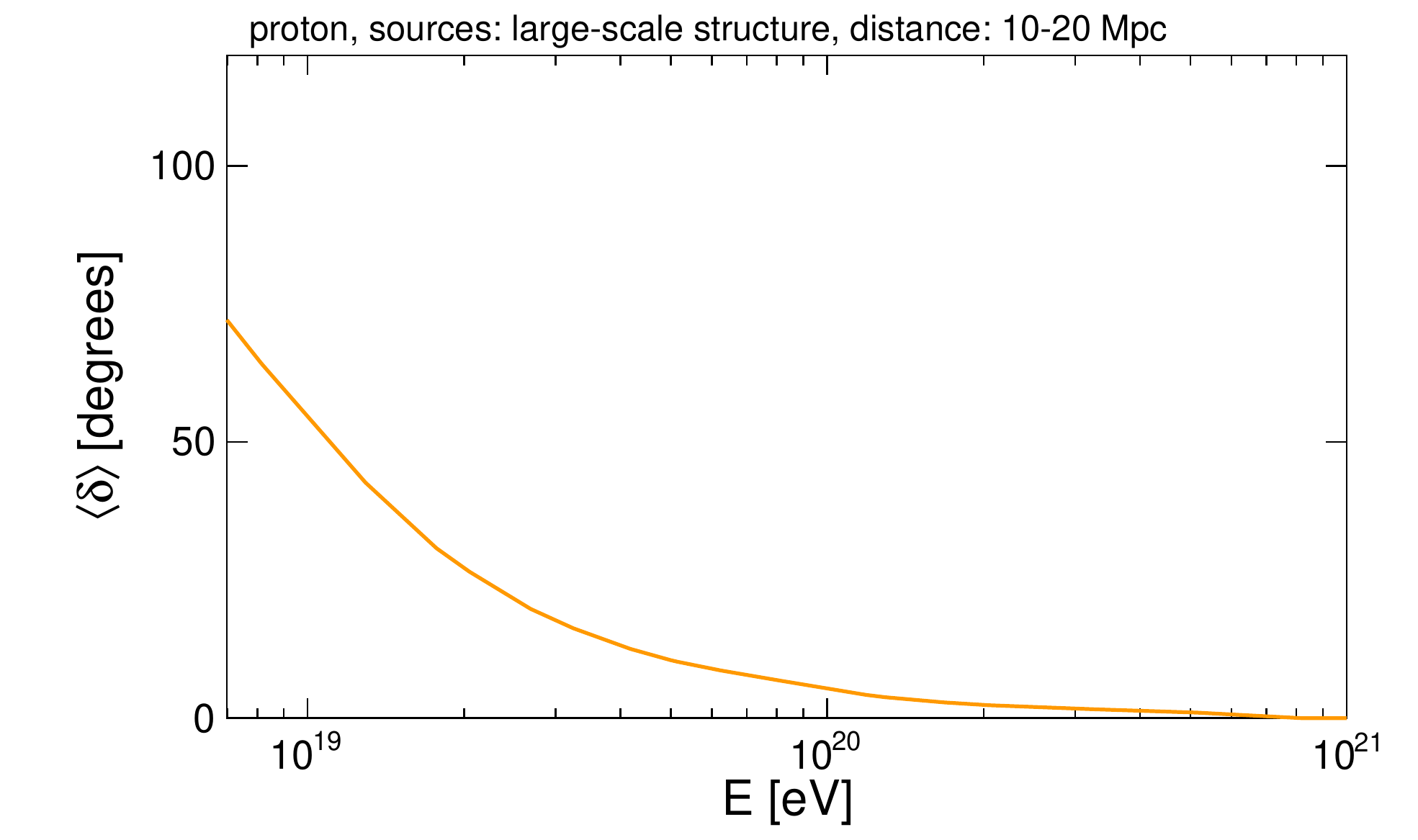}
	\caption{Average deflection as a function of the energy, and their corresponding standard deviations (hatched regions) for sources emitting protons with spectrum given by Eq.~\ref{eq:spec}. Sources are assumed to follow the simulated baryon density, and are distant 0-10 Mpc (left panel) and 10-20 Mpc (right panel). Spectral parameters are the same as in Fig.~\ref{fig:spec}.}
	\label{fig:deflE_Scaling_nearby_p}
\end{figure*}

We have not accounted for deflections in the GMF. One could, in principle, use the model by Jansson and Farrar~\cite{jansson2012a,jansson2012b} to estimate galactic deflections. This model is one of the most complete models of the GMF, but even so, it has its limitations~\cite{farrar2014a}. As shown in Ref.~\cite{jansson2012a}, for a proton with energy $E=60\;\text{EeV}$ the average deflection due to the GMF is $5.2^\circ$, with deflections larger than $2.2^\circ$ in about a 75\% of the sky. Nevertheless, one should keep in mind that there are a few lines of sight along which deflections are small. 


To quantify the magnitude of deflections from nearby sources we present Fig.~\ref{fig:deflE_Scaling_nearby_p}, which is similar to Fig.~\ref{fig:deflE_Scaling_p} but for sources located up to 10 Mpc and between 10 and 20 Mpc from Earth. In this case, as expected, deflections are on average smaller, except for run O. Some representative deflections for different distance bins are summarised in table~\ref{tab:defl} for both proton and iron primaries.

\begin{table*}
	\caption{Average deflections and corresponding $1\sigma$ standard deviations of UHECRs from nearby sources in runs F, L, S, and O. Results are presented for the case of sources following the large scale structure, both for proton (p) and iron (Fe) primaries.}
	\begin{tabular}{ccccccc}
	\hline \hline
	nucleus & E [EeV] & D [Mpc]  & run F & run L & run S & run O \\
	\hline
 p & 10 &  0-10  &  $41.3^\circ \pm 16.2^\circ$ &  $60.0^\circ \pm 21.1^\circ$ &  $42.9^\circ \pm 17.4^\circ$ &  $81.6^\circ \pm 21.0^\circ$ \\
 p & 60 &   0-5  &  $ 4.3^\circ \pm  2.0^\circ$ &  $ 8.8^\circ \pm  6.1^\circ$ &  $ 4.7^\circ \pm  2.4^\circ$ &  $27.1^\circ \pm 16.0^\circ$ \\
 p & 60 &  0-10  &  $ 4.9^\circ \pm  2.4^\circ$ &  $ 9.8^\circ \pm  6.3^\circ$ &  $ 5.5^\circ \pm  2.7^\circ$ &  $27.3^\circ \pm 15.6^\circ$ \\
 p & 60 & 10-20  &  $ 7.4^\circ \pm  2.8^\circ$ &  $13.8^\circ \pm  7.9^\circ$ &  $ 7.2^\circ \pm  2.8^\circ$ &  $35.7^\circ \pm 15.3^\circ$ \\
 p & 60 & 20-30  &  $ 8.4^\circ \pm  3.3^\circ$ &  $15.3^\circ \pm  7.0^\circ$ &  $ 9.0^\circ \pm  3.5^\circ$ &  $44.0^\circ \pm 16.8^\circ$ \\
p & 100 &   0-5  &  $ 2.3^\circ \pm  1.1^\circ$ &  $ 5.0^\circ \pm  3.2^\circ$ &  $ 2.7^\circ \pm  1.4^\circ$ &  $14.7^\circ \pm 11.5^\circ$ \\
p & 100 &  0-10  &  $ 2.8^\circ \pm  1.3^\circ$ &  $ 5.6^\circ \pm  3.4^\circ$ &  $ 2.9^\circ \pm  1.5^\circ$ &  $16.0^\circ \pm 10.9^\circ$ \\
p & 100 & 10-20  &  $ 4.1^\circ \pm  1.7^\circ$ &  $ 8.5^\circ \pm  4.9^\circ$ &  $ 4.3^\circ \pm  1.8^\circ$ &  $23.9^\circ \pm 12.3^\circ$ \\
p & 100 & 20-30  &  $ 4.8^\circ \pm  1.7^\circ$ &  $ 8.9^\circ \pm  3.2^\circ$ &  $ 5.0^\circ \pm  1.6^\circ$ &  $26.6^\circ \pm 12.4^\circ$ \\
Fe & 60 &   0-5  &  $76.6^\circ \pm 23.5^\circ$ &  $82.0^\circ \pm 20.6^\circ$ &  $70.0^\circ \pm 21.8^\circ$ &  $77.0^\circ \pm 19.5^\circ$ \\
Fe & 60 &  0-10  &  $78.6^\circ \pm 22.4^\circ$ &  $83.0^\circ \pm 20.2^\circ$ &  $78.9^\circ \pm 21.5^\circ$ &  $81.2^\circ \pm 19.4^\circ$ \\
Fe & 100 &   0-5 &  $56.0^\circ \pm 20.6^\circ$ &  $68.1^\circ \pm 21.1^\circ$ &  $60.9^\circ \pm 21.5^\circ$ &  $75.5^\circ \pm 24.1^\circ$ \\
Fe & 100 &  0-10 &  $64.0^\circ \pm 20.3^\circ$ &  $71.2^\circ \pm 21.4^\circ$ &  $64.8^\circ \pm 21.6^\circ$ &  $73.6^\circ \pm 22.1^\circ$ \\
	\hline \hline
	\end{tabular}
	\label{tab:defl}
\end{table*}

As can be seen in Figs.~\ref{fig:deflE_Scaling_nearby_p} and Tab.~\ref{tab:defl}, for a void field $\sim 1 \; \text{nG}$, for $E \approx 10 \; \text{EeV}$ deflections are large ($\gtrsim 20^\circ$), even if sources are closer than 10 Mpc. At 50 EeV deflections start to decrease, and for runs F and S, considering the variances, one expects a few lines of sight along which deflections are relatively small ($\lesssim 5^\circ$) if sources are distant less than 10 Mpc from Earth. At 100 EeV deflections are, in general, smaller than $\sim 5^\circ$ for $D<10\;\text{Mpc}$, except for run O. In this same energy range, for $10 \; \text{Mpc} < D < 20 \; \text{Mpc} $, runs O and L have high deflections ($\gtrsim 20^\circ$).


To first order, UHECR deflections are
\begin{equation}
	\delta \approx 0.9^\circ Z \left( \frac{100 \; \text{EeV}}{E} \right) \sqrt{\frac{l_c}{\text{Mpc}}} \sqrt{\frac{D}{10 \; \text{Mpc}}} \left( \frac{B}{\text{nG}} \right),
	\label{eq:defl}
\end{equation}
thus implying that for a source located at a distance $D$ emitting particles with rigidity $E/Z$, the deflection will grow as $B \sqrt{l_c}$. As can be seen in Fig.~\ref{fig:Bseed}, the power on large scales (small $k$) for run O is much larger than for runs F and S, i.e. $(B\sqrt{l_c})_O > (B\sqrt{l_c})_F > (B\sqrt{l_c})_S$, which explains the behaviours observed in Figs.~\ref{fig:deflE_Scaling_p} and~\ref{fig:deflE_Scaling_nearby_p}.

\begin{figure*}
	\includegraphics[width=\columnwidth]{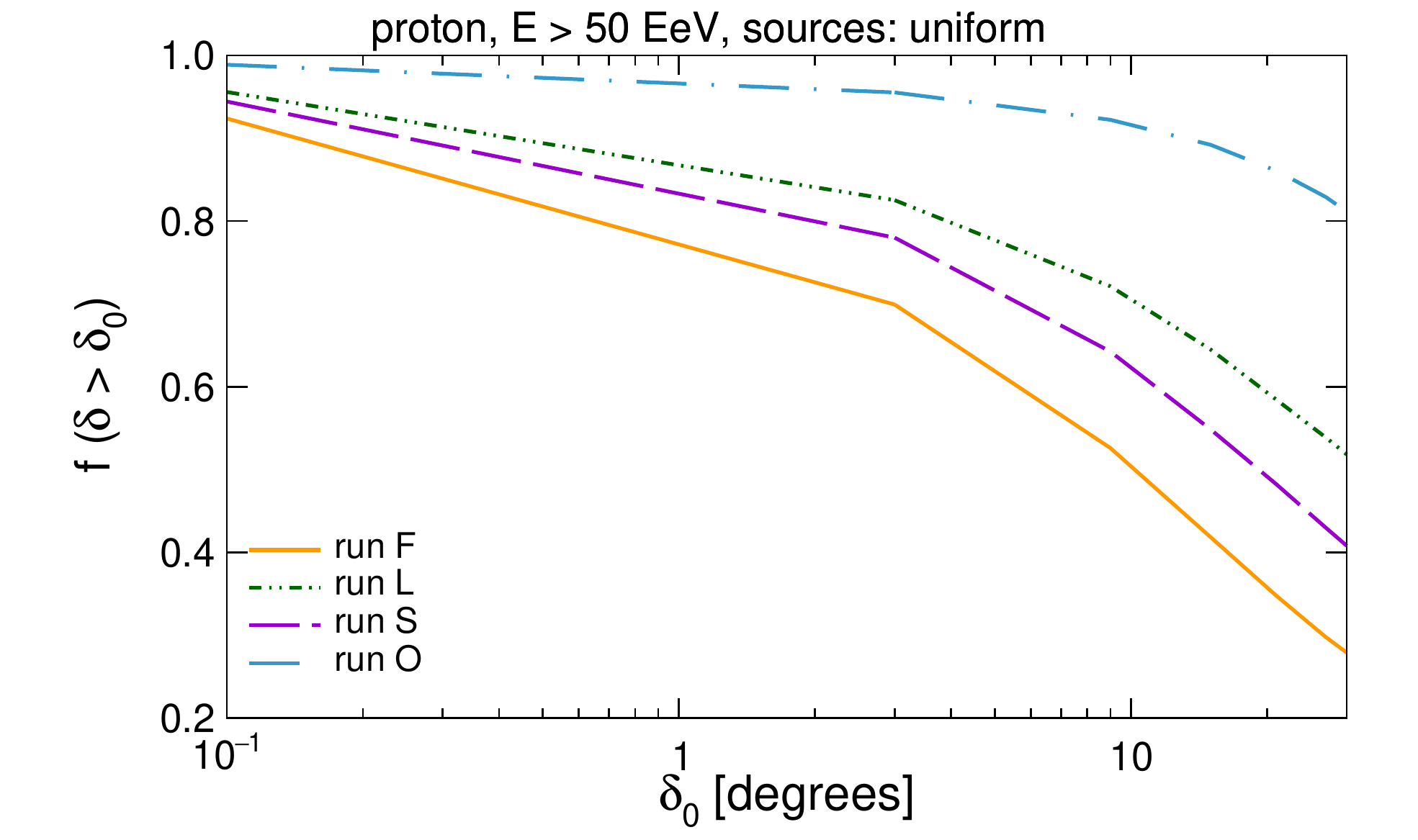}
	\includegraphics[width=\columnwidth]{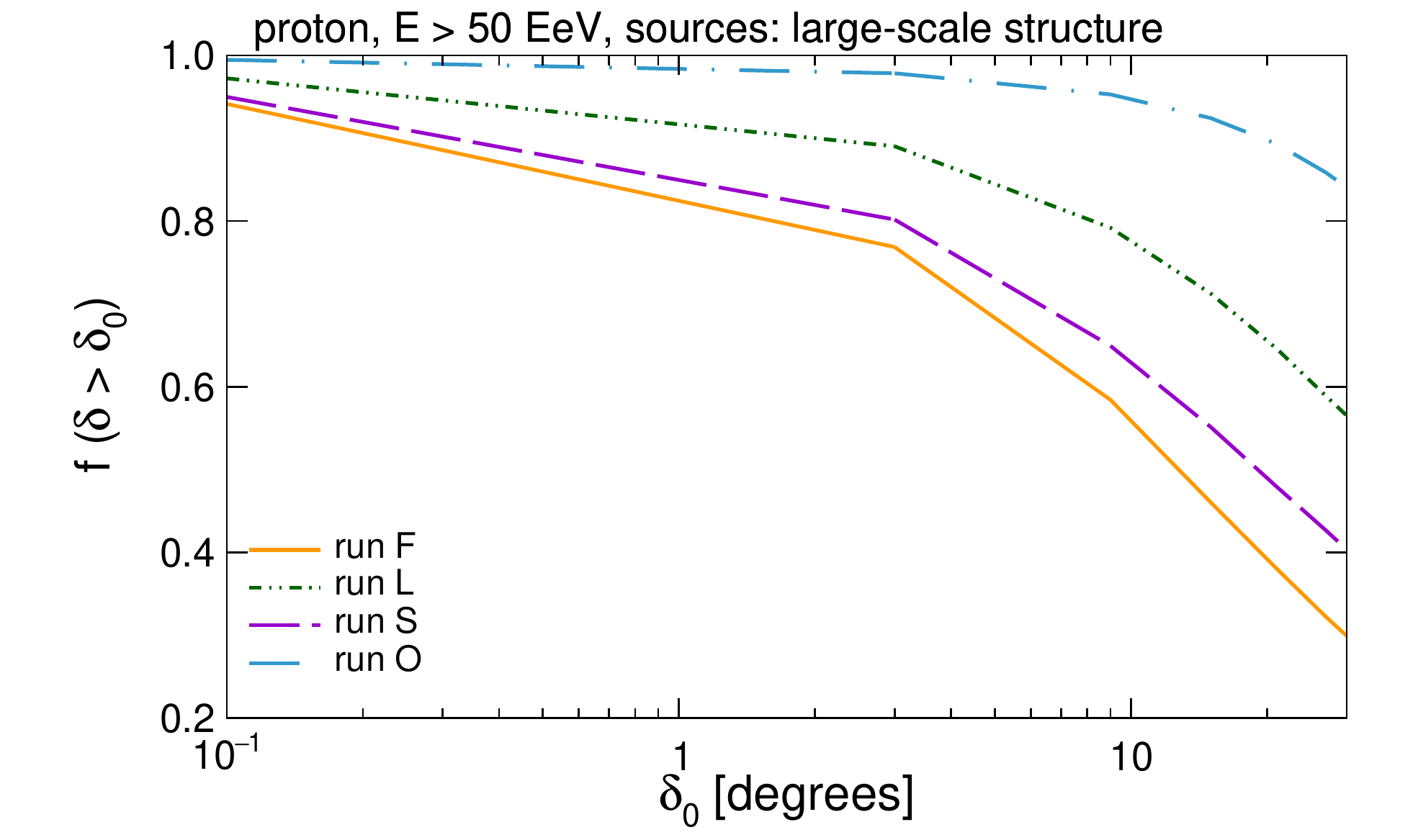}
	\caption{Fraction ($f$) of the sky with deflections ($\delta$) larger than a given value ($\delta_0$). Sources are assumed to be emitting protons with spectrum given by Eq.~\ref{eq:spec}, and either to follow the simulated baryon density (right) or to be uniformly distributed (left panel). The energy range considered here is $E > 50 \; \text{EeV}$. Spectral parameters are the same as in Fig.~\ref{fig:spec}.}
	\label{fig:cumDefl}
\end{figure*}

Another way to infer the impact of different magnetic field configurations on UHECR deflections is to estimate the fraction ($f$) of the sky with deflections larger than a given reference value ($\delta_0$). This is shown in Fig.~\ref{fig:cumDefl}.

From Fig.~\ref{fig:cumDefl} one can estimate the fraction of the sky with deflections smaller than a given threshold value, say $\delta_0 = 5^\circ$, which reads 0.25, 0.12, 0.22, 0.03, for grids F, L, S, and O, respectivelly, assuming that sources follow the baryon density. 

Therefore, uncertainties in the power spectrum of primordial fields may have a large impact on UHECR deflection estimates. Because the exact shape of the power spectrum is intrinsically related to the magnetogenesis mechanism that gave rise to primeval fields, which is unknown, this constitutes another source of uncertainty in the modelling of UHECR deflections. 


\subsection{Auger and TA hotspots}
 
 Many studies~\cite{rieger2009a,kachelriess2009a,yueksel2012a,farrar2013a,keivani2015a} have considered the possibility of Centaurus A (Cen A), the nearest active galactic nucleus, to be a source of UHECRs. Cen A is distant 3.7 Mpc from Earth, so one would expect extragalactic deflections to be small. In fact, Auger has observed an excess of UHECRs with energies $E > 58 \; \text{EeV}$ within a window of 15$^\circ$ around Cen A~\cite{auger2015b}. The (penalised) probability of such excess arising by chance from an isotropic distribution is 1.4\%. 

As shown in Ref.~\cite{farrar2013a}, adopting the Janson-Farrar model for the GMF~\cite{jansson2012a,jansson2012b}, UHE protons from Cen A would be deflected by approximately $3.8^\circ \pm 0.3^\circ$. 
Our results suggest that, even if we take into account deflections in the GMF, we could observe UHE protons from Cen A for $E \gtrsim  60  \; \text{EeV}$ for all magnetic field configurations considered. Inhomogeneities in the source distribution, however, makes deflections in the case of run O hard to be predicted. If the composition is predominantly iron nuclei, deflections of 100 EeV cosmic rays from Cen A would be large, as shown in Tab.~\ref{tab:defl}. 


One should note that due to the proximity of Cen A to Earth, only detailed models of both the galactic and extragalactic magnetic fields would allow us to make stronger claims, as opposed to our scenario which is an average over many observers.

TA has found evidences of an intermediate-scale anisotropy with statistical significance $5.1\sigma$ (pre-trial) and $3.4\sigma$ (post-trial). The region of about $20^\circ$ around $(l,b)=(177.4^\circ, 50.2^\circ)$ is close to the Ursa Major cluster, approximately 20 Mpc away from Earth. As can be seen in Tab.~\ref{tab:defl}, for sources distant 20-30 Mpc from Earth, deflections of 60 EeV protons are of the order of $10^\circ$, except for run O. Therefore, for the scenario we have considered, in which voids have nG fields, the Ursa Major cluster is a viable explanation for the TA hotspot if the composition is proton-dominated. For iron nuclei, however, deflections are much larger and this hypothesis would be disfavoured.

\subsection{Magnetisation near the observer}

\begin{figure*}
	\centering
	\includegraphics[width=\columnwidth]{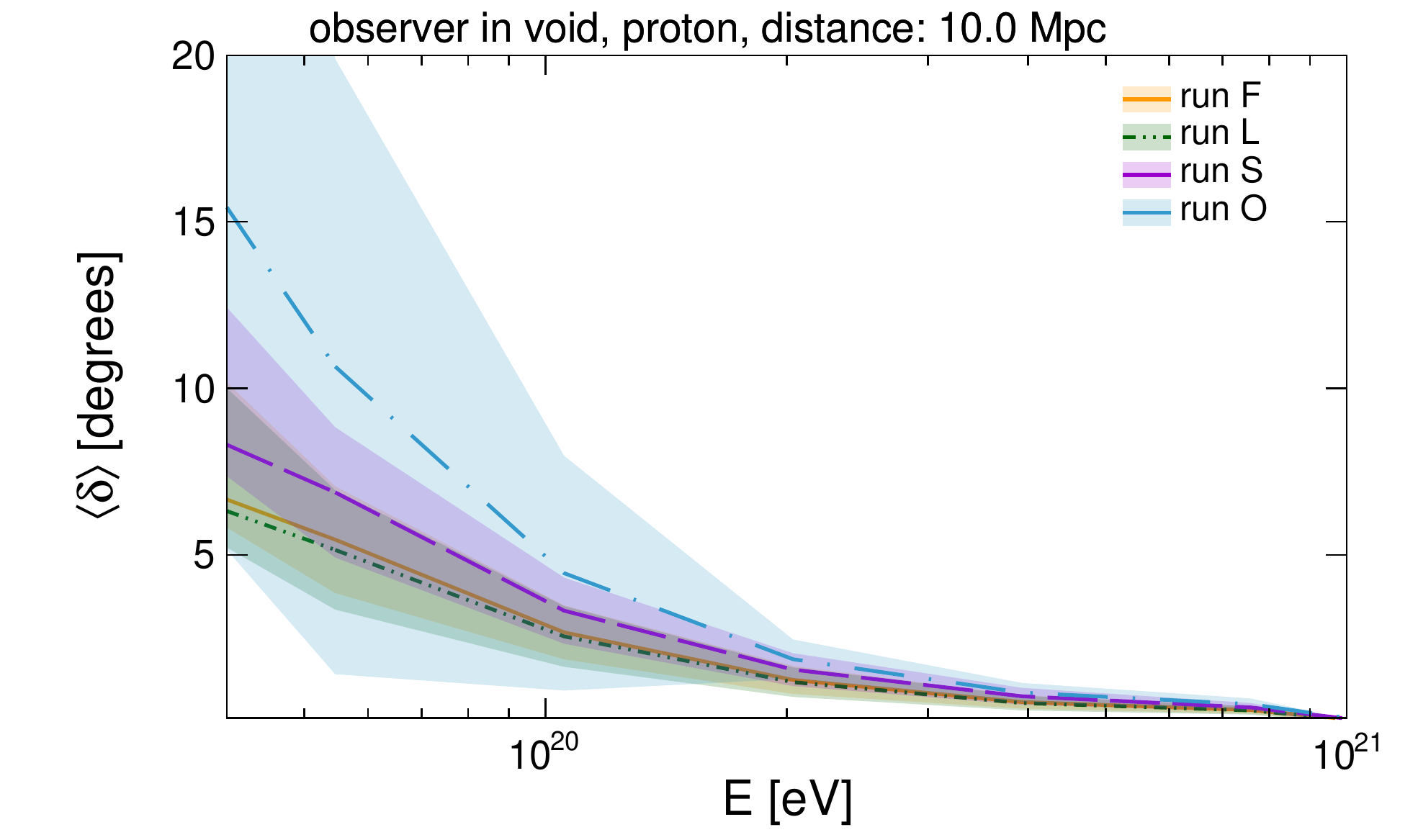}
	\includegraphics[width=\columnwidth]{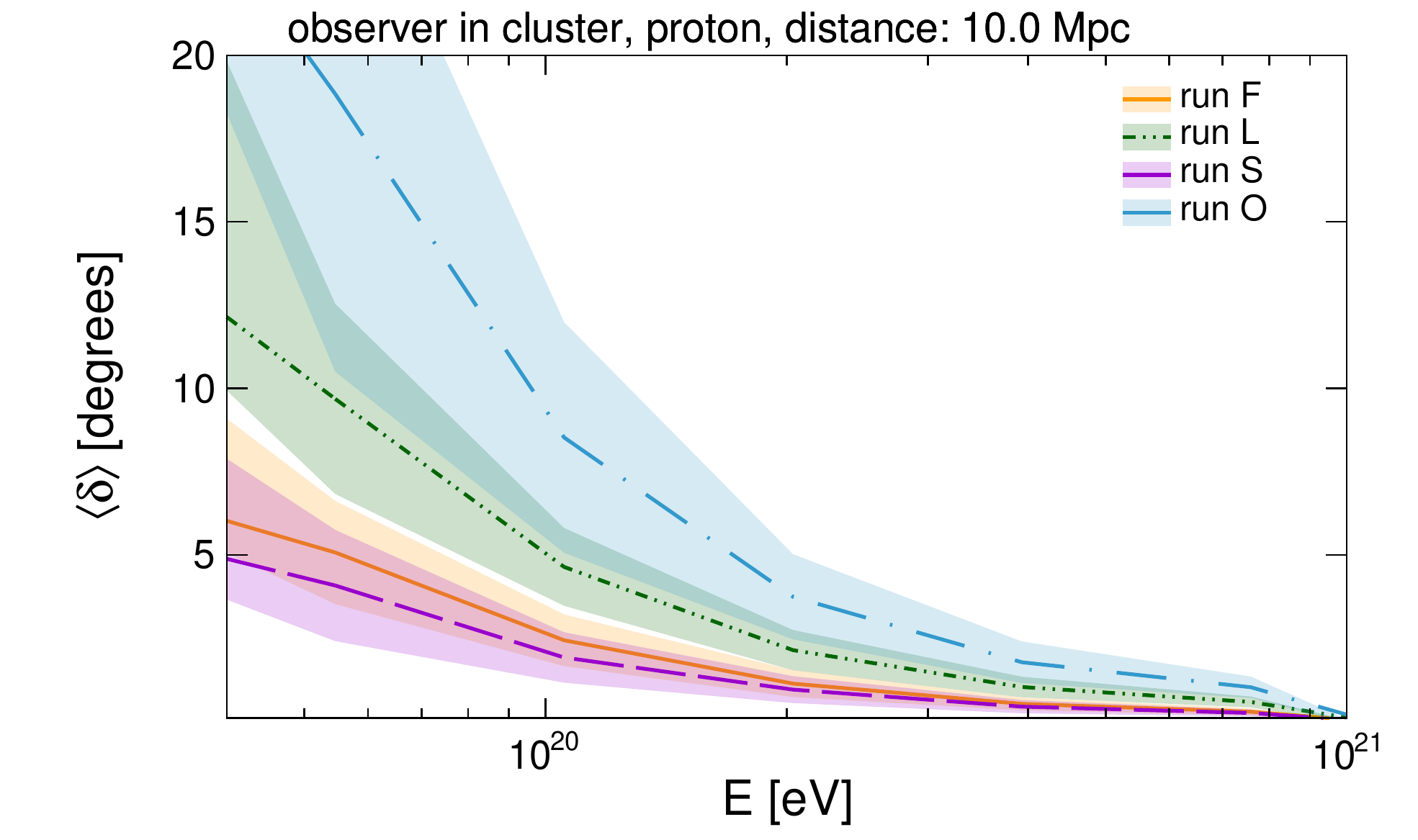}
	\caption{Average deflection as a function of the energy, and their corresponding standard deviations (hatched regions) for an observer in a void (left) and in a cluster (right panel), assuming that sources following the baryon density of the cosmic web are emitting protons with spectrum given by Eq.~\ref{eq:spec}. Deflections are calculated at a distance of 10 Mpc. Spectral parameters are the same as in Fig.~\ref{fig:spec}.}
	\label{fig:deflE_observers}
\end{figure*}

If both the observer and the source are immersed in highly magnetised regions, and are relatively close together (in the same filament, for instance), the average magnetic field effectively contributing for UHECR deflections is evidently much higher than in the case of an observer and a source separated by low magnetic field regions (e.g. voids). In our approach deflections are averaged over many observers to account for cosmic variance. However, it is important to assess the impact of different observer positions on the results, something that is not completely captured by the standard deviations from Tab.~\ref{tab:defl} and Figs.~\ref{fig:deflE_Scaling_p} and~\ref{fig:deflE_Scaling_nearby_p}. To this end, we select two observers. The first one is located close to the centre of a galaxy cluster; the other is in the centre of a void. The average deflections as a function of the energy at a distance 10 Mpc from the observer  are shown in Fig.~\ref{fig:deflE_observers}.

As one can see in Fig.~\ref{fig:deflE_observers}, the prospects for detecting UHE protons depend on the magnetisation near the observer. However, at energies of 100 EeV, a magnetic field configuration other that of run O would favour cosmic-ray astronomy with protons, as deflections would be of the order of a few degrees for sources distant up to 50 Mpc. In the case of iron, deflections exceed $10^\circ$ even for sources as close as 5 Mpc. Although not explicitly estimated, deflections of intermediate-mass nuclei such as nitrogen would likely allow us to identify a few nearby sources, especially for magnetic fields such as in run S, provided that the magnetisation near the observer is not too high.

\section{Summary and Outlook}\label{sec:summary}

We have analysed large-scale structure simulations which include magnetic fields based on four different configurations for the power spectra of the seed magnetic fields. We normalise the fields in the voids to Planck's upper bound of $\sim 1 \; \text{nG}$. These analyses provide valuable insights on horizons of UHECRs and on the actual impact of uncertainties related to extragalactic magnetic fields on cosmic-ray observables. 

Even for the maximally allowed magnetic fields, protons from nearby sources will not be deflected more than 15$^\circ$ in at least 25\% of the sky, except for run O, in which case this fraction is $\simeq 10\%$. This opens the possibility for UHE proton astronomy. Our results will not be changed qualitatively by considering additional deflection in the GMF for proton primaries with $E > 50 \; \text{EeV}$, since deflections are expected to be $\lesssim 2^\circ$ in about 25\% of the sky~\cite{jansson2012a}.

We have shown that observables such as spectrum and composition depend on the configuration of the magnetic field. In the scenarios studied herein, these uncertainties can be as high as $\simeq 80\%$ for the spectrum, depending on its normalisation, but do not exceed 5\% for the composition observables ($\langle \ln A \rangle$ and $\sigma(\ln A)$). This can potentially affect combined spectrum-composition fits such as the one recently done by the Pierre Auger Collaboration~\cite{auger2017a}.

In our MHD simulations voids retain information of the primordial magnetic power spectrum. Nevertheless, there are many processes unaccounted for that may pollute them. Chief amongst these processes are stellar and AGN feedbacks, which can eject magnetised material into the intergalactic medium in time scales comparable to the age of the universe, thus effectively erasing possible imprints of primordial fields. Because these were neglected in this work, our conclusion that UHECRs are sensitive to the primordial magnetic power spectrum may not hold. On the other hand, if the feedback is small, UHECRs could be useful probes of primordial fields.

We have shown that UHECR deflections tend to be dominated by the fields in the voids. Nevertheless, due to the many uncertainties involved, it would be a difficult task to use UHECRs to constrain the power spectrum of cosmic magnetic fields, albeit not impossible. For instance, if we have a few lines of sight between Earth and known sources, then the morphology of the arrival directions of the cosmic rays might allow us to infer some properties of the intervening magnetic field.

In summary, we have shown that in the most extreme case of voids having $\sim \; \text{nG}$ fields UHECR astronomy might be possible  depending on the power spectrum of magnetic fields and on the composition of the arriving cosmic rays. We have shown that in a significant portion of the sky UHECRs could be detected. Because a fraction of the total UHECR flux may be composed of protons, the ability to infer the composition on an event-by-event basis would be important to unambiguously identify the sources of UHECRs and should be considered when planning the next generation of UHECR experiments. 

We have studied the case in which intergalactic magnetic fields are close to their upper limit. In reality, they may be much weaker, which could enable the identification of sources even in the iron scenario. Ultimately, the ability to pinpoint the elusive sources of the highest energy cosmic rays depends not only the composition of cosmic rays and the distribution of sources, but also on the power spectrum of magnetic fields.

\section*{Acknowledgements}

RAB acknowledges the support of the Balzan Foundation via the Balzan Fellowship in early stages of this work. GS is supported by the DFG through collaborative research centre SFB 676, by the Helmholtz Alliance for Astroparticle Physics (HAP) funded by the Initiative and Networking Fund of the Helmholtz Association and by the Bundesministerium f\"ur Bildung und Forschung (BMBF).

\bibliography{references}

\end{document}